\documentclass[12pt, a4paper]{article}
\usepackage{amssymb}
\usepackage{amsmath}
\usepackage[dvips]{graphicx}
\usepackage{slashed}

\begin{document}

\baselineskip=18pt   
\numberwithin{equation}{section}
\allowdisplaybreaks
\pagestyle{myheadings}
\thispagestyle{empty}
\pagestyle{plain}

\def\eg{{\it e.g.}}
\newcommand{\nc}{\newcommand}
\nc{\rnc}{\renewcommand}
\rnc{\d}{\mathrm{d}}
\nc{\D}{\partial}
\nc{\K}{\kappa}
\nc{\bK}{\bar{\K}}
\nc{\bN}{\bar{N}}
\nc{\bq}{\bar{q}}
\nc{\vbq}{\vec{\bar{q}}}
\nc{\g}{\gamma}
\nc{\lrarrow}{\leftrightarrow}
\nc{\rg}{\sqrt{g}}
\rnc{\[}{\begin{equation}}
\rnc{\]}{\end{equation}}
\nc{\bea}{\begin{eqnarray}}
\nc{\eea}{\end{eqnarray}}
\nc{\nn}{\nonumber}
\rnc{\(}{\left(}
\rnc{\)}{\right)}
\nc{\q}{\vec{q}}
\nc{\x}{\vec{x}}
\rnc{\a}{a}  
\rnc{\b}{b}  
\nc{\ep}{\epsilon}
\nc{\tto}{\rightarrow}
\rnc{\inf}{\infty}
\rnc{\Re}{\mathrm{Re}}
\rnc{\Im}{\mathrm{Im}}
\nc{\z}{\zeta}
\nc{\mA}{\mathcal{A}}
\nc{\mB}{\mathcal{B}}
\nc{\mC}{\mathcal{C}}
\nc{\mD}{\mathcal{D}}
\nc{\mE}{\mathcal{E}}
\nc{\mF}{\mathcal{F}}
\rnc{\H}{\mathcal{H}}
\rnc{\L}{\mathcal{L}}
\nc{\<}{\langle}
\rnc{\>}{\rangle}
\nc{\fnl}{f_{NL}}
\nc{\fnleq}{f_{NL}^{equil.}}
\nc{\fnlloc}{f_{NL}^{local}}
\nc{\vphi}{\varphi}
\nc{\Lie}{\pounds}
\nc{\half}{\frac{1}{2}}
\nc{\bOmega}{\bar{\Omega}}
\nc{\bLambda}{\bar{\Lambda}}
\nc{\dN}{\delta N}
\nc{\gYM}{g_{\mathrm{YM}}}
\nc{\geff}{g_{\mathrm{eff}}}
\nc{\bg}{\hat{\gamma}}
\nc{\Oi}{\Omega_{[2]}}
\nc{\Oii}{\Omega_{[3]}}
\nc{\Ei}{E_{[2]}}
\nc{\Eii}{E_{[3]}}
\nc{\bOi}{\bar{\Omega}_{[2]}}
\nc{\bOii}{\bar{\Omega}_{[3]}}
\nc{\bEi}{\bar{E}_{[2]}}
\nc{\bEii}{\bar{E}_{[3]}}

\begin{titlepage}

\begin{center}

\hfill {\small ITFA-11-08}

\vskip 2 cm {\Large \bf Cosmological 3-point correlators from holography}
\vskip 1.25 cm {\bf Paul McFadden${}^1$ and Kostas Skenderis$\,{}^{1,2,3}$}
\\ {\vskip 0.5cm \it\small
${}^1\,$Institute for Theoretical Physics,
${}^2\,$Gravitation and Astro-Particle Physics Amsterdam, \\
${}^3\,$Korteweg-de$\,$Vries Institute for Mathematics, \\
Science Park 904, 1090 GL Amsterdam, the Netherlands.}

{\vskip 0.2cm \small
{\it E-mail:} {\tt P.L.McFadden@uva.nl, K.Skenderis@uva.nl} }

\end{center}

\vskip 1 cm

\begin{abstract}
\baselineskip=16pt

We investigate the non-Gaussianity of primordial cosmological perturbations using holographic methods.
In particular, we derive holographic formulae that relate all cosmological 3-point correlation functions, including both scalar and tensor perturbations, to stress-energy correlation functions of a holographically dual three-dimensional quantum field theory. These results apply to general single scalar inflationary universes that at late times approach either de Sitter spacetime or accelerating power-law cosmologies. We further show that in Einstein gravity all 3-point functions involving tensors may be obtained from correlators containing only positive helicity gravitons, with the ratios of these to 
the correlators involving one negative helicity graviton being given by universal functions of momenta, irrespectively of the potential of the scalar field.

As a by-product of this investigation, we obtain holographic formulae for the full 3-point function of the stress-energy tensor along general holographic RG flows.  These results should have applications in a wider holographic context.

\end{abstract}

\end{titlepage}

{\baselineskip=18pt
\setcounter{tocdepth}{2}
\tableofcontents
}

\section{Introduction}

This is the companion paper to \cite{McFadden:2010vh}. In
\cite{McFadden:2010vh}, we discussed the holographic computation of the
3-point function of scalar perturbations, and in this paper we compute
the 3-point functions involving both scalar and tensor
perturbations. The principal motivation for this work is
theoretical. We would like to understand whether standard cosmological
3-point functions can be recast in a form that is consistent with an
underlying holographic duality.  Such a reformulation would provide
strong evidence for this putative duality, and furthermore,
irrespective of the existence of a dual theory, 
it would also bring in fresh intuition about the
cosmological formulae and allow for quantum field
theory results and techniques to be used in cosmological computations.

The first indication that such a reformulation is possible was provided in
\cite{Maldacena:2002vr} where it was shown that
standard results for cosmological observables in (near) de Sitter (dS) spacetimes may be obtained by a certain analytic continuation
from corresponding results in Anti-de Sitter (AdS) spacetime. In turn, AdS results may be related to CFT correlation functions
via the AdS/CFT correspondence. In the same paper, it was argued that the putative dual theory computes the wavefunction
of the (near) de Sitter universe. Further work along these lines may be found in \cite{Larsen:2002et,Larsen:2003pf,vanderSchaar:2003sz,Larsen:2004kf}.

These results are very suggestive but one may wonder whether they are generic, 
{\it i.e.}, to what extent do they depend on special properties of the dS background?
It is well known that de Sitter is related to Anti de Sitter by an analytic continuation that takes the de Sitter time $t$ to $i r$,
where $r$ is the AdS radial coordinate, and the dS radius $L_{dS}$ to $i L_{AdS}$, where $L_{AdS}$ is the AdS radius.
What if one considers a general FRW spacetime? Is there an analogue of the
analytic continuation between AdS and dS that one may use in order to establish a holographic dictionary?

It turns out that such an analogue does indeed exist: one can show that for every FRW solution of a model with potential $V$,
there is a corresponding domain-wall solution of a model with potential $-V$ \cite{Skenderis:2006fb, Cvetic:1994ya}. 
A special case of this
domain-wall/cosmology (DW/C) correspondence is the relation between dS and AdS. Moreover, inflationary spacetimes that at late times approach
either dS spacetime or accelerating power-law cosmologies are mapped to asymptotically AdS spacetimes, and spacetimes that asymptotically approach
the near-horizon limit of the non-conformal branes, respectively. In both cases there is an established holographic dictionary \cite{Skenderis:2002wp, Kanitscheider:2008kd}
(these backgrounds then represent holographic RG flows), 
and one may hope to use it in order to relate cosmological observables to correlation functions of a dual QFT.

One should emphasise that a correspondence between highly symmetric spacetimes (such as the correspondence between the homogeneous and isotropic FRW spacetimes and the domain-wall solutions)
does not in general guarantee that generic perturbations around them will also
be in correspondence, and moreover this correspondence may be violated at the quantum level.
For example, the Feynman propagators of massive scalar fields in AdS and dS spacetime do not map to
each other under the analytic continuation mentioned above, see for example \cite{Das:2006wg}.

We thus undertook the task of checking explicitly whether or not
cosmological observables, such as the power spectra and non-Gausianities for general single scalar inflationary universes, can be related to
correlation functions of a dual QFT. These correlation functions are obtained from the corresponding domain-wall (DW) spacetime using the standard gauge/gravity
duality rules. In \cite{McFadden:2009fg,McFadden:2010jw}  we established that indeed the scalar and tensor power spectra are related to the 2-point function of the dual stress-energy tensor, while in \cite{McFadden:2010vh} we showed that the bispectrum of scalar perturbations is related to the 3-point function of the trace of the dual stress-energy tensor. 
Here, we will complete this task by showing that all cosmological 3-point functions are related to stress-energy correlation functions.

At first sight it might appear difficult
to establish such a relation for generic inflationary backgrounds,
since the computation of power spectra and non-Gausianities boils down to solving certain differential equations and these
can be explicitly integrated only for very symmetric backgrounds. Similarly, the explicit computation of correlation functions along
holographic RG flows requires solving specific differential equations, and their explicit integration is only possible for special
backgrounds (see \cite{BFS1,BFS2} for examples). Our strategy was thus to set up both computations in 
a manner that makes it manifest that the corresponding differential equations and boundary conditions map to each other under the
correspondence. It follows that one may recast the standard cosmological observables in terms of (an analytic continuation) of
strongly coupled QFT correlators, even though one may not be able to explicitly compute them.

On the holography side, 
the objects that enter the computation are 2-
and 3-point functions of the stress-energy tensor along general
holographic RG flows. Here, by general holographic RG flows, we mean
domain-wall spacetimes that under gauge/gravity duality correspond 
either to QFTs that in the UV approach a fixed point (asymptotically AdS
domain-walls), or to QFTs with generalised conformal structure that run
due to the dimensionality of their coupling constant (domain-walls
that asymptotically approach the non-conformal brane backgrounds). The
stress-energy tensor 2-point functions for general holographic RG flows
were
discussed in \cite{Papadimitriou:2004rz} and in
\cite{Kanitscheider:2008kd} for the two respective cases, using the
radial Hamiltonian formalism developed in
\cite{Papadimitriou:2004ap}. In this formalism, the central object is
the radial canonical momentum, {\it i.e.}, the canonical momentum in a
canonical formalism where the radial coordinate plays the role of
time.  The holographic correlators are then related to the response
functions, which are the coefficients in the expansion of the radial
canonical momentum in terms of the perturbations. Schematically, if $\Pi$
is the radial canonical momentum conjugate to the fluctuation $\zeta$
then we write
\[
\Pi = \Omega_{[2]} \zeta + \Omega_{[3]} \zeta^2 + \cdots
\]
The response functions are $\Omega_{[2]}$ and $\Omega_{[3]}$,
and are related to the 2- and 3-point function respectively 
of the operator dual to $\zeta$.
Here, we extend this formalism to encompass the 3-point function of the stress-energy
tensor, again for general holographic RG flows. These results should
thus have applications in a wider holographic context.

On the cosmology side, the objects of interest are the in-in tree-level
3-point functions. We computed these using a Hamiltonian
formalism and found that they also may be determined using 
response functions. The response functions in this case are defined as the
coefficients in the expansion of the standard canonical momentum in
terms of perturbations.  Furthermore, the DW/C correspondence maps the
domain-wall response functions to their cosmological analogues by a
simple analytic continuation.  Combined with the holographic
computation described in the previous paragraph, this provides a
holographic dictionary that relates cosmological observables to
correlation functions of a dual QFT.

We have thus shown by direct computation that, provided the standard
gauge/gravity duality holds, the cosmological spectrum and bispectrum have a
holographic interpretation: they are related to the analytic
continuation of 2- and 3-point functions of the stress-energy tensor
of a dual QFT.
Note that the DW/C correspondence can be expressed in terms
of an operation performed directly on the QFT: one analytically continues
the momenta and the rank of the gauge group. Thus, the QFT dual to the
inflationary spacetime is defined operationally by first performing computations
with the QFT dual to the DW spacetime, and then analytically continuing the
result. Previous discussions for a dual theory to dS spacetime
may be found in \cite{Witten:2001kn,Strominger:2001pn}.
Our results are also consistent with (but also
independent of) the interpretation of the
duality relation as providing the wavefunction of the universe 
\cite{Maldacena:2002vr}.

These holographic results were derived by working in the regime where the tree-level gravity
approximation is valid,
{\it i.e.}, the curvatures are small everywhere and gravity loops are suppressed.
A logical possibility is that this holographic interpretation
holds only in the regime in which it was derived.
On the other hand, the operational definition of the dual QFT
makes sense at least in large-$N$ perturbation theory
({\it i.e.}, we first  take the large-$N$ limit and then analytically continue) and
for any value of the (effective) 't Hooft coupling constant. One may then
use weakly coupled QFTs in the large-$N$ limit
in order to obtain novel scenarios for the
very early universe. In such scenarios, the universe started in a non-geometric
strongly coupled phase which is best described holographically
using the weakly coupled QFT. This leads to an interesting 
phenomenology \cite{McFadden:2010na}, and despite the fact that the
predictions of the holographic models differ from those of the empirical $\Lambda$CDM model
(and of generic single scalar slow-roll models),
they are still compatible with current data \cite{Dias:2011in,Easther:2011wh}.
In fact, a custom fit of the
WMAP and other astronomical data reveals that these models
are statistically comparable to $\Lambda$CDM.
In \cite{McFadden:2010vh}, we presented
the results for the non-Gaussianities of scalar perturbations
in such  scenarios. Interestingly, these are the only known models 
that lead to an exactly equilateral-type non-Gaussianity,
with $f_{NL}=5/36$ independently of all
parameters of the models (unfortunately, however, this value is
too small to be measured by
Planck). The corresponding results for
the non-Gaussianities we discuss here will be
presented elsewhere \cite{toappear}.

This paper is organised as follows. In the following section, 
we discuss domain-wall and cosmological spacetimes and their perturbations, 
and introduce the response functions. 
In Section \ref{cosmo_calcs} we compute the cosmological 3-point functions
in terms of response functions.  We also show in this section that for inflationary 
models based on Einstein gravity all 3-point functions involving 
tensors are determined from the correlators with only positive helicity gravitons.
In Section \ref{sec:Hol} we compute the holographic 3-point functions of 
the stress-energy tensor along general holographic RG flows. 
Section \ref{summary} contains the main results of this paper:
the holographic formulae that express the cosmological 3-point 
functions in terms of stress-energy tensor correlation functions.
Readers not interested in the derivation of these formulae may 
skip Sections \ref{cosmo_calcs} and \ref{sec:Hol} and proceed directly to this
section. In Section \ref{sec:discussion} we conclude. There are number of 
appendices: in Appendix \ref{App_GT}, we present the gauge-invariant perturbation
variables at quadratic order; in Appendix \ref{App_int}, we collect the 
cubic interaction terms; in Appendix \ref{App_helicity}, we discuss the 
helicity tensors; in Appendix \ref{App_Conventions}, we collect 
various conventions we use throughout the main text; in Appendix 
\ref{App_constraints}, we present the constraint equations at quadratic order,
and Appendix \ref{App_hol} contains the detailed derivation of the 
holographic results.

As this paper was finalised, \cite{Maldacena:2011nz} appeared containing
a related but complementary discussion of tensor non-Gaussianities.

\section{Perturbed domain-walls and cosmologies}

\subsection{Defining the perturbations}

Domain-walls and cosmologies may be described in a unified fashion via the ADM metric
\[
 \d s^2 = \sigma N^2 \d z^2 + g_{ij}(\d x^i+N^i\d z)(\d x^j+N^j\d z),
\]
where the perturbed lapse and shift functions may be written to second order as
\[
 N=1+\dN(z,\vec{x}), \qquad N_i=g_{ij}N^j=\dN_i(z,\vec{x}), \qquad g_{ij}= a^2(z)(\delta_{ij}+h_{ij}(z,\vec{x})),
\]
with $\sigma=+1$ for a Euclidean domain-wall
(whereupon $z$ becomes the transverse radial coordinate) and $\sigma=-1$ for a cosmology (whereupon $z$ becomes the cosmological proper time).
Taking the domain-wall to be Euclidean is convenient since the QFT vacuum implicit in the Euclidean formulation maps to the Bunch-Davies vacuum on the cosmology side, as discussed in \cite{McFadden:2010na}.
The spatial indices $i,j$ run from $1$ to $3$, and we have assumed (for simplicity) the background geometry to be spatially flat.

The $\delta g_{00}$ metric perturbation is then
\[
\label{00metric_perts}
 \delta g_{00} = 2\sigma \phi = \sigma (2\dN+\dN^2)+a^{-2} \dN_i \dN_i,
\]
where here, and in the remainder of the paper, we adopt the convention that repeated covariant indices are summed using the Kronecker delta
(in contrast, an index is raised or lowered by the full metric).
The remaining perturbations may be decomposed into scalar, vector and tensor pieces according to
\[
\label{metric_perts}
 \dN_i = a^2(\nu_{,i} + \nu_i), \qquad h_{ij} = -2\psi\delta_{ij}+2\chi_{,ij}+2\omega_{(i,j)}+\g_{ij},
\]
where the vector perturbations $\nu_i$ and $\omega_i$ are transverse, and the tensor perturbation $\g_{ij}$ is transverse traceless.
We similarly decompose the inflaton $\Phi$ into a background piece $\vphi$ and a perturbation $\delta\vphi$,
\[
 \Phi(z,\vec{x}) = \vphi(z)+\delta\vphi(z,\vec{x}).
\]
These formulae are understood to hold to second order in perturbation theory.

\subsection{Gauge-invariant variables}

We will work with the gauge-invariant variables $\z(z,\vec{x})$ and $\bg_{ij}(z,\vec{x})$, where $\z$ is the curvature perturbation on uniform energy density slices and
$\bg_{ij}$ is a transverse traceless tensor ($\bg_{ii}=0$ and $\D_i\bg_{ij}=0$).
These variables are defined such that in comoving gauge, where the inflaton perturbation $\delta\vphi$ vanishes, the spatial part of the perturbed metric reads
\[
g_{ij} = a^2 e^{2\z}[e^{\bg}]_{ij}=a^2 e^{2\z}(\delta_{ij}+\bg_{ij}+\frac{1}{2}\bg_{ik}\bg_{kj}).
\]
This implies the following general gauge-invariant definitions (see Appendix \ref{App_GT} for details)
\begin{align}
\label{zeta_gi}
\z &= -\psi-\frac{H}{\dot{\vphi}}\delta\vphi -\psi^2+\Big(\dot{H}-\frac{H\ddot{\vphi}}{\dot{\vphi}}\Big)\frac{\delta\vphi^2}{2\dot{\vphi}^2}
+\frac{H}{\dot{\vphi}^2}\delta\vphi\delta\dot{\vphi}+\frac{H}{\dot{\vphi}}(\chi_{,k}+\omega_k)\delta\vphi_{,k} + \frac{1}{4}\pi_{ij}X_{ij}, \\
\label{gb_gi}
\bg_{ij} &= \g_{ij} + \Pi_{ijkl}X_{kl},
\end{align}
where
\begin{align}
X_{ij} &= \frac{\sigma}{a^2\dot{\vphi}^2}\delta\vphi_{,i}\delta\vphi_{,j}-\frac{2}{a^2\dot{\vphi}}\dN_i\delta\vphi_{,j}
-\frac{\delta\vphi}{\dot{\vphi}}\dot{h}_{ij} - 2(\chi_{,k}+\omega_{k})_{,i}h_{jk} - (\chi_{,k}+\omega_k)h_{ij,k}
\nn\\
&\qquad + (\chi_{,k}+\omega_k)_{,i}(\chi_{,k}+\omega_k)_{,j} +2\psi\g_{ij} - \frac{1}{2}\g_{ik}\g_{kj},
\end{align}
and the transverse and transverse traceless projection operators $\pi_{ij}$ and $\Pi_{ijkl}$ are defined as
\[
\label{projection_operators}
 \pi_{ij} = \delta_{ij}-\frac{\D_i\D_j}{\D^2}, \qquad \Pi_{ijkl}=\frac{1}{2}\big(\pi_{ik}\pi_{jl}+\pi_{il}\pi_{jk}-\pi_{ij}\pi_{kl}\big).
\]
Here, and throughout, we use dots to denote differentiation with respect to $z$, and we define $H=\dot{a}/a$ and $\ep=-\dot{H}/H^2$.

\subsection{Equations of motion} 

Our action comprises a single scalar field minimally coupled to gravity with a potential $V(\Phi)$.
In the ADM formalism, the combined domain-wall/cosmology action takes the form
\[
\label{FullLagrangian}
S= \frac{1}{2\K^2}\int\d^4x N\sqrt{g} \left[K_{ij}K^{ij}-K^2+N^{-2}(\dot{\Phi}-N^i\Phi_{,i})^2
+\sigma \(-R+g^{ij}\Phi_{,i}\Phi_{,j}+2\K^2 V(\Phi)\)\right],
\]
where $\kappa^2=8\pi G$ and $K_{ij}=[(1/2)\dot{g}_{ij}-\nabla_{(i}N_{j)}]/N$ is the extrinsic curvature of constant-$z$ slices.
In this expression, spatial gradients and potential terms appear with positive sign for Euclidean domain-walls and with negative sign for Lorentzian cosmologies as expected.

We will restrict our consideration to background solutions in which the evolution of the scalar field $\vphi(z)$ is
(piece-wise) monotonic in $z$.
For such solutions, $\vphi(z)$ can be inverted to $z(\vphi)$, allowing $H$ to be re-expressed as a function of $\vphi$, {\it i.e.}, $H(z) = -(1/2)W(\vphi)$.  The complete equations of motion for the background then take the simple first-order form
\[
\label{bgd_eom}
\frac{\dot{a}}{a} =-\frac{1}{2}W, \qquad \dot{\vphi}= W_{,\vphi}, \qquad 2\sigma\K^2V = (W_{,\vphi})^2-\frac{3}{2}W^2.
\]

Turning now to the perturbations, one may derive an action for the gauge-invariant fluctuations $\z$ and $\bg_{ij}$ by
solving the Hamiltonian and momentum constraints and backsubstituting into the Lagrangian $\mathcal{L}$, as described in \cite{Maldacena:2002vr}.
To compute 3-point functions, we will need this action to cubic order, keeping careful track of the sign $\sigma$.
The full result may be found in Appendix \ref{App_int}.

To connect with the holographic analysis in later sections, however, it is most convenient to describe the perturbations in the Hamiltonian formalism.  To this end, we define the quantities
\[
 \Pi = \frac{\D(\K^2\mathcal{L})}{\D \dot{\z}}, \qquad \Pi_{ij} = \frac{\D(\K^2\mathcal{L})}{\D\dot{\bg}_{ij}},
\]
corresponding to ($\K^2$ times) the canonical momenta with respect to $\z$ and with respect to $\bg_{ij}$.
When working in momentum space, it is useful to decompose the transverse traceless tensors $\Pi_{ij}$ and $\bg_{ij}$ in a helicity basis as
\[
\bg_{ij}(\q) = \bg^{(s)}(\q)\ep^{(s)}_{ij}(\q) , \qquad  \Pi_{ij}(\q) = \Pi^{(s)}(\q)\ep^{(s)}_{ij}(\q),
\]
where here, and throughout, we assume the summation of repeated helicity indices over the values $\pm1$.
Our conventions for the helicity tensors $\ep^{(s)}_{ij}(\q)$ are summarised in Appendix \ref{App_helicity}.

The full Hamiltonian may then be written
\[
 H = H^{(2)}+H^{(3)}, \qquad H^{(3)} = H_{\z\z\z}+H_{\z\z\bg}+H_{\z\bg\bg}+H_{\bg\bg\bg}
\]
where the free part
\begin{align}
\label{H_free}
 \K^2 H^{(2)} = \int [\d q] \Big[ &
\frac{1}{4a^3\ep}\Pi(\q)\Pi(-\q)
+\frac{4}{a^3}\Pi^{(s)}(\q)\Pi^{(s)}(-\q) \nn\\
& -\sigma a \ep q^2\z(\q)\z(-\q)
-\frac{\sigma a}{4}q^2\bg^{(s)}(\q)\bg^{(s)}(-\q)\Big].
\end{align}
The bracket notation we use here (and throughout) for the various measures appearing in momentum space integrals is described in Appendix \ref{App_Conventions}.
The interaction term $H_{\z\z\z}$ may be found in \cite{McFadden:2010vh}, however we will have no use for it here.
The remaining pieces of the cubic interaction Hamiltonian then take the form
\begin{align}
\label{H_zzg}
\K^2 H_{\z\z\bg} &= \int [[\d q_1\d q_2 \d q_3]] \Big[ \mA^{(s_3)}_{123} \z(-\q_1)\z(-\q_2)\bg^{(s_3)}(-\q_3) + \mB^{(s_3)}_{123}\z(-\q_1)\z(-\q_2)\Pi^{(s_3)}(-\q_3) \nn \\
&\qquad + \mC^{(s_3)}_{123}\z(-\q_1)\Pi(-\q_2)\bg^{(s_3)}(-\q_3)+\mD^{(s_3)}_{123}\z(-\q_1)\Pi(-\q_2)\Pi^{(s_3)}(-\q_3) \nn \\
&\qquad + \mE^{(s_3)}_{123}\Pi(-\q_1)\Pi(-\q_2)\bg^{(s_3)}(-\q_3)+\mF^{(s_3)}_{123}\Pi(-\q_1)\Pi(-\q_2)\Pi^{(s_3)}(-\q_3)\Big], \\[2ex]
\label{H_zgg}
\K^2 H_{\z\bg\bg} &= \int [[\d q_1\d q_2 \d q_3]] \Big[\mA^{(s_2 s_3)}_{123} \z(-\q_1)\bg^{(s_2)}(-\q_2)\bg^{(s_3)}(-\q_3) \nn\\
&\qquad + \mB^{(s_2 s_3)}_{123}\z(-\q_1)\bg^{(s_2)}(-\q_2)\Pi^{(s_3)}(-\q_3)
+ \mC^{(s_2 s_3)}_{123}\z(-\q_1)\Pi^{(s_2)}(-\q_2)\Pi^{(s_3)}(-\q_3) \nn\\
&\qquad +\mD^{(s_2 s_3)}_{123}\Pi(-\q_1)\bg^{(s_2)}(-\q_2)\bg^{(s_3)}(-\q_3)
+ \mE^{(s_2 s_3)}_{123}\Pi(-\q_1)\bg^{(s_2)}(-\q_2)\Pi^{(s_3)}(-\q_3) \nn\\
&\qquad +\mF^{(s_2 s_3)}_{123}\Pi(-\q_1)\Pi^{(s_2)}(-\q_2)\Pi^{(s_3)}(-\q_3)\Big], \\[2ex]
\label{H_ggg}
\K^2 H_{\bg\bg\bg} &= \int [[\d q_1 \d q_2 \d q_3]] \mA^{(s_1s_2s_3)}_{123}\bg^{(s_1)}(-\q_1)\bg^{(s_2)}(-\q_2)\bg^{(s_3)}(-\q_3),
\end{align}
where the coefficients are appropriately symmetrised functions of the helicities $s_i$ and the magnitudes of the momenta, which we denote
$q_i = +\sqrt{\q_i^{\,2}}$.
The precise form of these coefficients is given in Appendix \ref{App_int}.

In the following section, we will make frequent use of Hamilton's equations, which read
\begin{align}
\label{Hamilton_eqs}
&\dot{\z}(\q) = (2\pi)^3 \frac{\D(\K^2 H)}{\D \Pi(-\q)}, \qquad
&&\dot{\bg}^{(s)}(\q) = \half\, (2\pi)^3 \frac{\D (\K^2 H)}{\D \Pi^{(s)}(-\q)}, \nn\\[2ex]
&\dot{\Pi}(\q) = -(2\pi)^3\frac{\D(\K^2 H)}{\D \z(-\q)}, \qquad
&&\dot{\Pi}^{(s)}(\q) = -\half\,(2\pi)^3\frac{\D(\K^2 H)}{\D\bg^{(s)}(-\q)} .
\end{align}
Note in particular the factors of one half multiplying the r.h.s.~of the equations for $\dot{\bg}^{(s)}$ and $\dot{\Pi}^{(s)}$.
These factors arise from the standard normalisation convention for helicity tensors (see Appendix \ref{App_helicity}).

\subsection{Response functions}

Given a perturbative solution of the classical equations of motion for $\z$ and $\bg^{(s)}$, we may formally expand the associated canonical momenta
$\Pi$ and $\Pi^{(s)}$ in terms of $\z$ and $\bg^{(s)}$ to any given order in perturbation theory.
At quadratic order, we may thus write
\begin{align}
\label{response_fns1}
 \Pi(\q_1) &= \Omega_{[2]}(q_1)\z(\q_1)+\int[[\d q_2\d q_3]]\,\Big[\Omega_{[3]}(q_i)\z(-\q_2)\z(-\q_3) \nn\\
&\qquad\qquad +
\Omega^{(s_3)}_{[3]}(q_i)\z(-\q_2)\bg^{(s_3)}(-\q_3) +\Omega^{(s_2 s_3)}_{[3]}(q_i)\bg^{(s_2)}(-\q_2)\bg^{(s_3)}(-\q_3)\Big] \\[2ex]
\label{response_fns2}
\Pi^{(s_1)}(\q_1) &= E_{[2]}(q_1)\bg^{(s_1)}(\q_1) + \int[[\d q_2\d q_3]]\,\Big[ {\Eii^{(s_1)}}(q_i)\z(-\q_2)\z(-\q_3)
\nn\\
&\qquad\qquad
+ \Eii^{(s_1 s_3)}(q_i)\z(-\q_2)\bg^{(s_3)}(-\q_3) +\Eii^{(s_1 s_2 s_3)}(q_i)\bg^{(s_2)}(-\q_2)\bg^{(s_3)}(-\q_3)\Big],
\end{align}
where our notation for the integration measure in these formulae is explained in Appendix \ref{App_Conventions}.
We will refer to the various functions $\Omega$ and $E$ defined by these equations as {\it response functions}.

Hamilton's equations \eqref{Hamilton_eqs} imply the linear response functions $\Oi(q)$ and $\Ei(q)$ satisfy
\[
\label{Linear_response_eom}
0 = \dot{\Omega}_{[2]}(q)+\frac{1}{2a^3\ep}\Omega_{[2]}^2(q)- 2\sigma a\ep q^2, \qquad
0 = \dot{E}_{[2]}(q)+\frac{4}{a^3}E^2_{[2]}(q)-\frac{\sigma a}{4}q^2.
\]
Given solutions $\z_q$ and $\bg_q$ of the linearised equations of motion (obeying Bunch-Davies vacuum condition at early times),
\[
\label{linear_eom}
0=\ddot{\z}_q+(3H+\dot{\ep}/\ep)\dot{\z}_q-\sigma a^{-2} q^2 \z_q, \qquad
0=\ddot{\bg}_q+3H\dot{\bg}_q-\sigma a^{-2} q^2\bg_q,
\]
we may then solve \eqref{Linear_response_eom} to find
\[
 \Oi(q) = 2a^3\ep\,\frac{ \dot{\z}_q}{\z_q}, \qquad \Ei(q) = \frac{a^3}{4}\,\frac{\dot{\bg}_q}{\bg_q}.
\]
Given these solutions for the linear response functions $\Oi(q)$ and $\Ei(q)$, we
may then solve for the response functions appearing at quadratic order as follows.
For example, to find $\Omega^{(s)}_{[3]}$, we note that Hamilton's equations \eqref{Hamilton_eqs}, after making use of \eqref{response_fns1}, read
\begin{align}
\label{zdot}
 \dot{\z}(\q_1) &= \frac{1}{2a^3\ep}\Oi(\q_1)\z(\q_1)+\int[[\d q_2\d q_3]]\z(-\q_2)\bg^{(s_3)}(-\q_3)\Big[\frac{1}{2a^3\ep}\Oii^{(s_3)}(q_i)+ \mC^{(s_3)}_{213}  \nn\\
&\qquad\qquad\qquad\qquad\qquad +\mD^{(s_3)}_{213} \Ei(q_3) +2\mE^{(s_3)}_{123}\Oi(q_2)+2\mF^{(s_3)}_{123}\Oi(q_2)\Ei(q_3)\Big] +\ldots,\\[1ex]
\label{pidot}
\dot{\Pi}(\q_1) &= 2\sigma a \ep q_1^2\z(\q_1)-\int[[\d q_2\d q_3]] \z(-\q_2)\bg^{(s_3)}(-\q_3) \Big[2\mA^{(s_3)}_{123} +2\mB^{(s_3)}_{123}\Ei(q_3)
+ \mC^{(s_3)}_{123}\Oi(q_2) \nn\\
&\qquad\qquad\qquad\qquad\qquad\qquad\qquad\qquad\qquad +\mD^{(s_3)}_{123}\Oi(q_2)\Ei(q_3)\Big] +\ldots,
\end{align}
where we have retained only quadratic terms of the form $\z\bg$.
On the other hand, differentiating \eqref{response_fns1} directly, we have
\begin{align}
 \dot{\Pi}(\q_1) &= \dot{\Omega}_{[2]}(q_1)\z(\q_1)+\Oi(q_1)\dot{\z}(\q_1)+\int[[\d q_2\d q_3]]\Big[
\dot{\Omega}^{(s_3)}_{[3]}(q_i)\z(-\q_2)\bg^{(s_3)}(-\q_3) \nn\\
&\qquad\qquad + \Oii^{(s_3)}(q_i)\Big(\dot{\z}(-\q_2)\bg^{(s_3)}(-\q_3)+\z(-\q_2)\dot{\bg}^{(s_3)}(-\q_3)\Big)\Big]
+\ldots
\end{align}
Using Hamilton's equations to replace $\dot{\z}$ and $\dot{\bg}^{(s)}$, comparing with \eqref{pidot} we then find
\begin{align}
\label{Omega3s_eom}
0 &= \dot{\Omega}^{(s_3)}_{[3]}(q_i) +\Big[\frac{1}{2a^3\ep}\big(\Omega_{[2]}(q_1)+\Omega_{[2]}(q_2)\big)+\frac{4}{a^3}E_{[2]}(q_3)\Big]
\Omega^{(s_3)}_{[3]}(q_i) + \mathcal{X}^{(s_3)}_{123},
\end{align}
where
\begin{align}
\label{Xs}
\mathcal{X}^{(s_3)}_{123} &= 2\mA^{(s_3)}_{123}+2\mB^{(s_3)}_{123}E_{[2]}(q_3)+\mC^{(s_3)}_{123}\Omega_{[2]}(q_2)+\mC^{(s_3)}_{213}\Omega_{[2]}(q_1) +\mD^{(s_3)}_{123}\Oi(q_2)\Ei(q_3) \nn \\
&\quad +\mD^{(s_3)}_{213}\Oi(q_1)\Ei(q_3)
+2\mE^{(s_3)}_{123}\Oi(q_1)\Oi(q_2)+2\mF^{(s_3)}_{123}\Oi(q_1)\Oi(q_2)\Ei(q_3).
\end{align}

Thus, given solutions $\z_q$ and $\bg_q$ of the linearised equations of motion
we may solve \eqref{Omega3s_eom} to find
\[
\label{Omega3s_soln}
 \Oii^{(s_3)}(z, q_i)= -\frac{1}{\z_{q_1}(z)\z_{q_2}(z)\bg_{q_3}(z)}\int_{z_0}^z\d z' \mathcal{X}^{(s_3)}_{123}(z')\z_{q_1}(z')\z_{q_2}(z')\bg_{q_3}(z').
\]
We will return to the choice of lower limit $z_0$ in this integral in the next subsection.

The remaining response functions may be obtained by an analogous procedure.  The response function $\Oii$ is derived in \cite{McFadden:2010vh} (where it is called $\Lambda$), however we will not need it here.  For the rest, one finds
\[
\label{Omega3ss_soln}
 \Oii^{(s_2 s_3)}(z, q_i)= -\frac{1}{\z_{q_1}(z)\bg_{q_2}(z)\bg_{q_3}(z)}\int_{z_0}^z\d z' \mathcal{X}^{(s_2s_3)}_{123}(z')\z_{q_1}(z')\bg_{q_2}(z')\bg_{q_3}(z').
\]
where
\begin{align}
\label{Xss}
\mathcal{X}^{(s_2 s_3)}_{123} &= \mA^{(s_2s_3)}_{123}+\half\mB^{(s_2s_3)}_{123}E_{[2]}(q_3)+\half\mB^{(s_3s_2)}_{132}E_{[2]}(q_2)+\mC^{(s_2s_3)}_{123}E_{[2]}(q_2) E_{[2]}(q_3)+\mD^{(s_2s_3)}_{123}\Oi(q_1) \nn \\
&\quad +\half\mE^{(s_2s_3)}_{123}\Oi(q_1)\Ei(q_3)+\half\mE^{(s_3s_2)}_{132}\Oi(q_1)\Ei(q_2)+\mF^{(s_2s_3)}_{123}\Oi(q_1)\Ei(q_2)\Ei(q_3).
\end{align}
Similarly,
\[
\label{E3sss_soln}
 \Eii^{(s_1s_2 s_3)}(z, q_i)= -\frac{1}{\bg_{q_1}(z)\bg_{q_2}(z)\bg_{q_3}(z)}\int_{z_0}^z\d z' \,\frac{3}{2}\mA^{(s_1s_2s_3)}(z',q_i)\bg_{q_1}(z')\bg_{q_2}(z')\bg_{q_3}(z').
\]
Finally, we find
\begin{align}
\label{cross_relations}
\Eii^{(s_1)}(q_1,q_2,q_3) = \frac{1}{4}\Oii^{(s_1)}(q_2,q_3,q_1), \qquad
\Eii^{(s_1s_3)}(q_1,q_2,q_3) = \Oii^{(s_3s_1)}(q_2,q_3,q_1).
\end{align}

As we will see in the next section, the imaginary parts of these response functions give the various cosmological 3-point functions we wish to compute.

\subsection{Domain-wall/cosmology correspondence}

Examining the background equations of motion \eqref{bgd_eom}, as well as the Hamiltonian for the perturbations (both the free part \eqref{H_free} and the interaction terms \eqref{zzg_coeffts}, \eqref{zgg_coeffts} and \eqref{ggg_coeffts}),
we see that a perturbed cosmological solution ({\it i.e.}, with $\sigma=-1$) expressed in terms of $\K^2$ and $\q_i$ analytically continues to a perturbed domain-wall solution ($\sigma=+1$) expressed in terms of the analytically continued variables $\bK^2$ and $\vec{\bq}_i$, where
\[
\label{cont}
 \bK^2= -\K^2, \qquad \bq_i = -i q_i.
\]
The first continuation serves to reverse the sign of the potential in \eqref{bgd_eom} (taking, for example, dS to AdS), while the second ensures that $q_i^2=-\bq_i^2$, which mimics the effect of changing the sign of $\sigma$ in the Hamiltonian for the perturbations.  
The choice of branch cut in the continuation of the magnitude $q_i$ 
is imposed on us by the necessity of mapping the cosmological Bunch-Davies vacuum behaviour 
to the domain-wall solution that decays smoothly in the interior,
as required for the computation of holographic correlation functions.

Turning now to the response functions, we see that if we define the response functions
appearing in \eqref{response_fns1} and \eqref{response_fns2} to be cosmological response functions with $\sigma=-1$, then the corresponding 
domain-wall response functions, which we will denote using a bar, are given by analytic continuation of the momenta.  For example,
\[
\label{response_cont}
 \bOmega_{[2]}(\bq) = \bOmega_{[2]}(-iq) = \Omega_{[2]}(q), \qquad \bOii^{(s_3)}(\bq_i)=\bOii^{(s_3)}(-iq_i)=\Oii^{(s_3)}(q_i), \quad \mathrm{etc}.
\]

In the remainder of this paper, we will use the unbarred variables $\K^2$, $q_i$ and unbarred response functions for performing cosmological calculations, and the barred variables $\bK^2$, $\bq_i$ and barred response functions for domain-wall calculations.  To analytically continue the results from domain-walls to cosmologies, and vice versa, we use \eqref{cont} and \eqref{response_cont}.

Finally, we note the analytic continuation  \eqref{cont}, when translated into QFT variables, reads
\[
\label{QFT_cont}
 \bN = -iN, \qquad \bq_i=-iq_i,
\]
where $\bN$ is the rank of the gauge group of the QFT dual to the domain-wall spacetime, and $N$ is the rank of the gauge group of the pseudo-QFT dual to the corresponding cosmology. 
Our choice of branch cut in the continuation of $\bN$ ensures that the dimensionless effective QFT coupling $g_{\mathrm{eff}}^2=g_{\mathrm{YM}}^2\bN/\bq = g_{\mathrm{YM}}^2 N/q$ is invariant under \eqref{QFT_cont}.

\section{Cosmological 3-point functions}
\label{cosmo_calcs}

\subsection{Calculation using response functions}

We begin by quantising the interaction picture fields $\z$ and $\bg^{(s)}$ such that
\[
\label{mode_decomp}
\z(z,\q)=\a(\q)\z_q(z)+\a^\dag(-\q)\z_q^*(z), \qquad
\bg^{(s)}(z,\q)=\b^{(s)}(\q)\bg_q(z)+\b^{(s)\dag}(-\q)\bg_q^*(z),
\]
where the creation and annihilation operators obey the usual commutation relations
\[
\label{CCRs}
[\a(\q),\a^\dag(\q\,')] = (2\pi)^3 \delta(\q-\q\,'), \qquad
[\b^{(s)}(\q), \b^{(s)\dag}(\q\,')]=(2\pi)^3\delta(\q-\q\,')\delta^{s s'},
\]
and the mode functions $\z_q(z)$ and $\bg_q$ are solutions of the linearized equation of motion \eqref{linear_eom}, with initial conditions specified by the Bunch-Davies vacuum condition.

The corresponding 2-point functions are
\[
\label{2ptdefs}
\<\!\<\z(z,q)\z(z,-q)\>\!\>  = |\z_q(z)|^2, \qquad
\<\!\<\bg^{(s)}(z,q)\bg^{(s')}(z,-q)\>\!\> = |\bg_q(z)|^2\,\delta^{s s'},
\]
where the double bracket notation we use for correlators is defined in Appendix \ref{App_Conventions}
and serves to suppress the appearance of delta functions associated with overall momentum conservation.
As was shown in \cite{McFadden:2009fg, McFadden:2010na}, these 2-point functions may be re-expressed in terms of the linear response functions: 
\[
\label{powerspecresults}
\<\!\<\z(z,q)\z(z,-q)\>\!\>  = \frac{-\K^2}{2\Im[\Oi(z,q)]}, \quad
\<\!\<\bg^{(s)}(z,q)\bg^{(s')}(z,-q)\>\!\> = \frac{-\K^2 \delta^{s s'}}{4\Im[\Ei(z,q)]}.
\]

At tree level, the 3-point function is given in the in-in formalism by the standard formula \cite{Maldacena:2002vr, Weinberg:2005vy},
{\it e.g.},
\[
\label{zzg}
\<\z(z,\q_1)\z(z,\q_2)\bg^{(s_3)}(z,\q_3)\> = -i \int^z_{z_0} \d z' \< [{:}\z(z,\q_1)\z(z,\q_2)\bg^{(s_3)}(z,\q_3){:}\,,\,{:}H_{\z\z\bg}(z'){:}]\>.
\]
Here, to ensure convergence, a suitable infinitesimal rotation of the contour of integration is understood.  The lower limit $z_0$ represents some very early time (corresponding to large and negative conformal times) at which the interactions are assumed to be switched on.  Note that both the operators appearing in the commutator in this formula are taken to be normal ordered as indicated.

Inserting the operator equivalent of \eqref{H_zzg} for $H_{\z\z\bg}$ in the above formula, we may now proceed to evaluate the commutator explicitly, noting that for the cubic terms in $H_{\z\z\bg}$ we may replace
\begin{align}
\label{Pi_exp}
\Pi(z,\q) & = \a(\q)\,\Oi(z,q)\z_q(z)+\a^\dag(-\q)\,\Oi^*(z,q)\z^*_q(z), \nn\\
\Pi^{(s)}(z,\q) &= \b^{(s)}(\q)\Ei(z,q)\bg_q(z)+\b^{(s)\dag}(-\q)\Ei^*(z,q)\bg^*_q(z).
\end{align}
In this manner, we find the full 3-point function
\begin{align}
\label{result1}
\frac{\<\!\<\z(z,q_1)\z(z,q_2)\bg^{(s_3)}(z,q_3)\>\!\>}{|\z_{q_1}(z)|^2 |\z_{q_2}(z)|^2 |\bg_{q_3}(z)|^2}
&= \Im \Big[\frac{1}{\z_{q_1}(z)\z_{q_2}(z)\bg_{q_3}(z)}\int^z_{z_0} \d z' (-2\K^2) \mathcal{X}^{(s_3)}_{123}(z') \z_{q_1}(z')\z_{q_2}(z')\bg_{q_3}(z')\Big] \nn \\[2ex]
& = \Im \left[2\K^{-2}\Oii^{(s_3)}(z,q_1,q_2,q_3)\right],
\end{align}
where $\mathcal{X}^{(s_3)}_{123}$ is defined in \eqref{Xs} and in the last line we have used \eqref{Omega3s_soln}. The lower limit of integration $z_0$ in \eqref{Omega3s_soln} should thus be identified with the lower limit $z_0$ in \eqref{zzg}.
Again, our double bracket notation for correlators is given in Appendix \ref{App_Conventions}.

Applying the same procedure for the remaining two cosmological correlators, we find
\begin{align}
\label{result2}
\frac{\<\!\<\z(z,q_1)\bg^{(s_2)}(z,q_2)\bg^{(s_3)}(z,q_3)\>\!\>}{|\z_{q_1}(z)|^2 |\bg_{q_2}(z)|^2 |\bg_{q_3}(z)|^2}
&= \Im \left[4\K^{-2}\Oii^{(s_2 s_3)}(z,q_1,q_2,q_3)\right], \\[2ex]
\label{result3}
\frac{\<\!\<\bg^{(s_1)}(z,q_1)\bg^{(s_2)}(z,q_2)\bg^{(s_3)}(z,q_3)\>\!\>}{|\bg_{q_1}(z)|^2 |\bg_{q_2}(z)|^2 |\bg_{q_3}(z)|^2}
& = \Im \left[8\K^{-2}\Eii^{(s_1s_2 s_3)}(z,q_1,q_2,q_3)\right].
\end{align}

Equations \eqref{result1}, \eqref{result2} and \eqref{result3} are the main results of this section.  Used in combination with \eqref{powerspecresults},
they allow us to re-express the cosmological 3-point functions in terms of the corresponding response functions.

\subsection{Helicity structure of cosmological 3-point correlators}

In this section, we discuss the most general possible helicity structure for cosmological 3-point correlators involving tensors.
Since we are principally interested in their late-time behaviour, we will suppress all $z$-dependence.

First of all, symmetry under permutations imposes that
\begin{align}
\label{noparity}
\<\!\<\z(q_1)\z(q_2)\bg^{(s_3)}(q_3)\>\!\> & =  \hat{A}_{(12)3}+s_3 \hat{B}_{(12)3}, \nn\\
\<\!\<\z(q_1)\bg^{(s_2)}(q_2)\bg^{(s_3)}(q_3)\>\!\> & =  \tilde{A}_{1(23)}+s_2\tilde{B}_{1(23)}+s_3 \tilde{B}_{1(23)}+s_2s_3 \tilde{C}_{1(23)}, \nn\\
\<\!\<\bg^{(s_1)}(q_1)\bg^{(s_2)}(q_2)\bg^{(s_3)}(q_3)\>\!\> & =  A_{(123)}+s_1 B_{1(23)}+s_2 B_{2(31)}+s_3B_{3(12)} \nn \\
&\quad +
s_1s_2C_{(12)3}+s_2s_3C_{(23)1}+s_3s_1C_{(31)2}+s_1s_2s_3D_{(123)},
\end{align}
where the coefficients are appropriately symmetrised functions of the momenta, {\it i.e.}, $B_{(12)3}\equiv B(q_1,q_2,q_3)=B(q_2,q_1,q_3)$, {\it etc}.
If the interactions are invariant under parity ($\vec{q_i}\tto -\vec{q_i}$), then the correlation functions are in addition invariant under reversing the sign of all helicities, {\it i.e.}, $s_i \tto -s_i$.
In this case, the helicity structure simplifies to
\begin{align}
\label{parity}
\<\!\<\z(q_1)\z(q_2)\bg^{(s_3)}(q_3)\>\!\> & =  \hat{A}_{(12)3} \nn\\
\<\!\<\z(q_1)\bg^{(s_2)}(q_2)\bg^{(s_3)}(q_3)\>\!\> & =  \tilde{A}_{1(23)}+s_2s_3 \tilde{C}_{1(23)}, \nn\\
\<\!\<\bg^{(s_1)}(q_1)\bg^{(s_2)}(q_2)\bg^{(s_3)}(q_3)\>\!\> & =  A_{(123)} + (s_1s_2C_{(12)3} + \mathrm{cyclic\,\,perms}).
\end{align}
These relations encode the observation that, for example, all correlators involving three tensors may be obtained from
either $\<\!\<\bg^{(+)}(q_1)\bg^{(+)}(q_2)\bg^{(+)}(q_3)\>\!\>$ or $\<\!\<\bg^{(+)}(q_1)\bg^{(+)}(q_2)\bg^{(-)}(q_3)\>\!\>$ through permutations and parity.

It is interesting to note that the helicity structure of 3-point correlators arising from the standard inflationary Lagrangian \eqref{FullLagrangian} takes the form
\begin{align}
\label{hel1}
 \<\!\<\z(q_1)\z(q_2)\bg^{(s_3)}(q_3)\>\!\> &= F_{\z\z\bg}(q_i) \theta^{(s_3)}(q_i), \\
\label{hel2}
\<\!\<\z(q_1)\bg^{(s_2)}(q_2)\bg^{(s_3)}(q_3)\>\!\> &= F_{\z\bg\bg}(q_i) \theta^{(s_2s_3)}(q_i), \\
\label{hel3}
\<\!\<\bg^{(s_1)}(q_1)\bg^{(s_2)}(q_2)\bg^{(s_3)}(q_3)\>\!\> &= F_{\bg\bg\bg}(q_i) \theta^{(s_1s_2s_3)}(q_i),
\end{align}
where $F_{\z\z\bg}(q_i)$, $F_{\z\bg\bg}(q_i)$ and $F_{\bg\bg\bg}(q_i)$ are general functions of the magnitudes $q_i$,
while $\theta^{(s_3)}(q_i)$,  $\theta^{(s_2s_3)}(q_i)$ and $\theta^{(s_1s_2s_3)}(q_i)$ denote specific contractions of helicity tensors given in Appendix
\ref{App_helicity}.
This structure follows simply from the observation that all terms in the cubic interaction Hamiltonians $\H_{\z\z\bg}$, $\H_{\z\bg\bg}$ and $\H_{\bg\bg\bg}$ are proportional to $\theta^{(s_3)}(q_i)$, $\theta^{(s_2s_3)}(q_i)$ and $\theta^{(s_1s_2s_3)}(q_i)$ respectively, 
as may be seen from \eqref{zzg_coeffts}, \eqref{zgg_coeffts} and \eqref{ggg_coeffts}.
Using the results of Appendix \ref{App_helicity}, we then find
\begin{align}
&\hat{A}_{(12)3}  = \frac{\lambda^2}{4\sqrt{2}\,q_3^2} F_{\z\z\bg}(q_i), &&\nn \\[1ex]
&\tilde{A}_{1(23)}  = \Big(1-\frac{\lambda^2}{8q_2^2q_3^2}\Big)F_{\z\bg\bg}(q_i),
\quad &&\tilde{C}_{1(23)} = \frac{1}{2q_2q_3}(-q_1^2+q_2^2+q_3^2)F_{\z\bg\bg}(q_i),
\nn\\[1ex]
& A_{(123)} =  \frac{\lambda^2}{4\sqrt{2}}\Big(\frac{1}{q_1^2}+\frac{1}{q_2^2}+\frac{1}{q_3^2}-\frac{\lambda^2}{8q_1^2q_2^2q_3^2}\Big) F_{\bg\bg\bg}(q_i),
\quad && C_{(12)3} =  \frac{\lambda^2(q_1^2+q_2^2+3q_3^2)}{8\sqrt{2}\,q_1 q_2 q_3^2}F_{\bg\bg\bg}(q_i),
\end{align}
where $\lambda$ is a function of the momenta given in \eqref{lambda_def}.
Equivalently, we have the following relationships between correlators:
\begin{align}
\label{consistency}
\<\!\<\z(q_1)\bg^{(+)}(q_2)\bg^{(+)}(q_3)\>\!\>
 &=
\left(\frac{(q_2+q_3)^2-q_1^2}{(q_2-q_3)^2-q_1^2}\right)^2 \<\!\<\z(q_1)\bg^{(+)}(q_2)\bg^{(-)}(q_3)\>\!\>,
\nn\\[2ex]
\<\!\<\bg^{(+)}(q_1)\bg^{(+)}(q_2)\bg^{(+)}(q_3)\>\!\>
&=
\left(\frac{q_1+q_2+q_3}{q_1+q_2-q_3}\right)^4 \<\!\<\bg^{(+)}(q_1)\bg^{(+)}(q_2)\bg^{(-)}(q_3)\>\!\>.
\end{align}
Note that these are universal relations that hold for any Lagrangian of the form \eqref{FullLagrangian}. In particular, they hold irrespectively of the form of the inflationary potential.
Using these relations one can reconstruct all 3-point correlators from those involving only positive helicity gravitons.

If one considers a more general Lagrangian in place of \eqref{FullLagrangian}, for example by including higher derivative interactions, then one must instead revert to the general form \eqref{parity} (or \eqref{noparity} if the interactions violate parity).
Note also that if the background possesses any isometries  ({\it e.g.}, the case where the background is exactly de Sitter spacetime), these may be used to constrain the form of the generalised shape functions appearing in these expressions.
This idea has been explored recently in \cite{Antoniadis:2011ib, Maldacena:2011nz}.

\subsection{An example: slow-roll inflation}

To illustrate the discussion above, in this subsection we compute the cosmological 3-point functions in the slow-roll approximation using response functions.  We will assume all momenta to be of comparable magnitude.

\subsubsection*{\it (i) Two scalars and a graviton}    

As shown by Maldacena \cite{Maldacena:2002vr}, the cubic action for two scalars and a graviton may be written to leading order in slow-roll as
\[
 \K^2 \L_{\z\z\bg} = a\ep\bg_{ij}\z_{,i}\z_{,j} + \ldots,
\]
after performing suitable field redefinitions (and setting $\sigma=-1$).  These field redefinitions may be neglected on super-horizon scales, however, and so can effectively be ignored in the following.
The interaction Hamiltonian then comprises only the single term
\[
 \mA^{(s_3)}_{123}=-a\ep\theta^{(s_3)}(q_i).
\]

To evaluate the response function $\Omega_{[3]}^{(s_3)}(q_i)$ at late times, one then uses \eqref{Omega3s_soln}, substituting in the de Sitter solutions
\[
 \z_q(\tau)\approx \frac{i\K H_*}{\sqrt{4\ep_* q^3}}(1+iq\tau)e^{-iq\tau}, \qquad
\bg_q(\tau)\approx \frac{i\K H_*}{\sqrt{q^3}}(1+iq\tau)e^{-iq\tau}.
\]
for the linearised mode functions.
Here, the asterisk indicates taking the values at the time of horizon crossing $z=z_*$ (where $q\approx a(z_*)H(z_*)$), while the conformal time $\tau = \int \d z/a$.
We find
\[
 \Oii^{(s_3)}(\tau,q_i) = \frac{2\ep_*}{H_*^2}\theta^{(s_3)}(q_i)\int^\tau_{-\inf}\frac{\d \tau'}{\tau'^2}(1+iq_1\tau')(1+iq_2\tau')(1+iq_3\tau')e^{-iq_t \tau'}
\]
where $q_t = \sum_i q_i$.  (Note that $a\approx -1/H_*\tau$ and time derivatives of $\ep_*$ and $H_*$ are higher order in slow roll).
While the full response function diverges as $\tau\tto 0^-$, the imaginary part is finite:
\[
\Im[\Omega_{[3]0}^{(s_3)}(q_i)] = \frac{2\ep_*}{H_*^2}\Big(q_t-\frac{\sum_{i<j}q_iq_j}{q_t}-\frac{q_1q_2q_3}{q_t^2}\Big)\,\theta^{(s_3)}(q_i),
\]
where the subscript zero indicates taking the value in the late-time limit.

From \eqref{result1}, we then obtain
\[
\<\!\<\z(q_1)\z(q_2)\bg^{(s_3)}(q_3)\>\!\> = \frac{\K^4H_*^4}{4\ep_* \prod_i q_i^3}\Big(q_t-\frac{\sum_{i<j}q_iq_j}{q_t}-\frac{q_1q_2q_3}{q_t^2}\Big)\,\theta^{(s_3)}(q_i),
\]
in agreement with \cite{Maldacena:2002vr}.

\subsubsection*{\it (ii) One scalar and two gravitons}

To leading order in slow-roll, the cubic action for one scalar and two gravitons may be written as \cite{Maldacena:2002vr}
\[
 \K^2 \L_{\z\bg\bg} = \frac{1}{2}a^5\ep H \dot{\bg}_{ij}\dot{\bg}_{ij} \partial^{-2}\dot{\z}_c +\ldots,
\]
where the field
\[
 \z_c = \z+\frac{1}{32}\bg_{ij}\bg_{ij}-\frac{1}{16}\partial^{-2}(\bg_{ij}\partial^2\bg_{ij})+\ldots
\]
Here, we have omitted further terms that may be neglected on superhorizon scales.
The cubic Hamiltonian for this sector is then once again only a single term:
\[
 \mF^{(s_2s_3)}_{123} = \frac{4H}{a^4 q_1^2}\theta^{(s_2s_3)}(q_i).
\]

We may evaluate the response function $\Oii^{(s_2s_3)}(q_i)$ using \eqref{Omega3ss_soln}, noting that the response functions $\Oi(q)$ and $\Ei(q)$ for the linearised fluctuations are
\[
\Oi(\tau,q) = \frac{2a^2\ep_*}{\z_q}\frac{\d\z_q}{\d\tau} \approx\frac{-2a\ep_*q^2}{H_*(1+iq\tau)}, \qquad
\Ei(\tau,q)=\frac{a^2}{4\bg_q}\frac{\d\bg_q}{\d\tau} \approx \frac{- a q^2}{4H_*(1+iq\tau)}.
\]
Consequently, taking the late-time limit, one finds using \eqref{result2} that
\[
 \<\!\<\z_c(q_1)\bg^{(s_2)}(q_2)\bg^{(s_3)}(q_3)\>\!\> = \frac{\K^4H_*^4}{2\prod_iq_i^3}\frac{q_2^2q_3^2}{q_t}\theta^{(s_2s_3)}(q_i).
\]

Reverting to the original $\z$ variable, we then find\footnote{We believe the coefficient of $k_1^3$ in equation (4.13) of \cite{Maldacena:2002vr} should be ${-}1/2$ rather than ${-}1/4$.}
\[
 \<\!\<\z(q_1)\bg^{(s_2)}(q_2)\bg^{(s_3)}(q_3)\>\!\> = \frac{\K^4H_*^4}{2\prod_iq_i^3}\Big(\frac{q_2^2q_3^2}{q_t}-\frac{1}{8}q_1(q_1^2-q_2^2-q_3^2)\Big)\,\theta^{(s_2s_3)}(q_i).
\]

\subsubsection*{\it (iii) Three gravitons} 

The cubic interaction Hamiltonian for three gravitons is simply that given in \eqref{ggg_coeffts}, with $\sigma=-1$.
As above, we may solve for the response function $\Eii^{(s_1s_2s_3)}(q_i)$ using \eqref{E3sss_soln}.
From \eqref{result3}, we then find that at late times
\begin{align}
 \<\!\<\bg^{(s_1)}(q_1)\bg^{(s_2)}(q_2)\bg^{(s_3)}(q_3)\>\!\> &= \frac{\K^4H_*^4}{2\prod_i q_i^3}\Big(q_t-\frac{\sum_{i<j}q_iq_j}{q_t}-\frac{q_1q_2q_3}{q_t^2}\Big)\,\theta^{(s_1s_2s_3)}(q_i),
\end{align}
in agreement with \cite{Maldacena:2002vr}.

\section{Holographic 3-point functions \label{sec:Hol}}

In this section we present our holographic calculation for the full 3-point function of the stress-energy tensor.
We begin with a careful identification of the 3-point correlators appearing when the 1-point function in the presence of sources is expanded to quadratic order.  We then proceed with the holographic analysis itself, first for the case of asymptotically AdS domain-walls, and secondly for the case of asymptotically power-law domain-walls.

\subsection{Correlation functions of the stress-energy tensor}

Correlation functions of the stress-energy tensor may be obtained by 
coupling the theory to a background metric $g_{(0)kl}$ and functionally 
differentiating with respect to the metric. Equivalently, 
starting with the 1-point function in the presence of sources,
$\<T_{ij}\>_s = (2/\sqrt{ g_{(0)}})\delta S/\delta g_{(0)}^{ij}$, 
higher correlation functions may be obtained through repeated
functional differentiation with respect to the source $g_{(0)kl}$,
after which the source is set to its background value.  In performing
this operation, one must be careful to note that the stress-energy
tensor $T_{ij}$ has itself a purely classical dependence on the
metric: this additional metric dependence gives rise to contact terms,
some of which we need to keep track of.  Specifically, when computing
the 3-point function, we need to retain {\it semi-local} contact terms
in which only two of the three points involved are coincident,
since terms of this form contribute to local-type non-Gaussianity.
We may, on the other hand, discard {\it ultralocal} contact terms in which all three points are coincident: such terms are generically scheme dependent ({\it i.e.}, one may remove them by addition of finite local counterterms).

Expanding the 1-point function in the presence of sources to quadratic order about a flat background, we have
\[
 \delta\<T^i_j(\x_1)\>_s = \big(\delta^{ip}+\delta g^{ip}(\x_1)\big)\delta\<T_{pj}(\x_1)\>_s
\]
(as the vacuum expectation value $\<T_{pj}(\x_1)\>$ vanishes),
where
\[
\delta\<T_{pj}(\x_1)\>_s = \int \d^3\x_2 \frac{\delta\<T_{pj}(\x_1)\>}{\delta g^{kl}(\x_2)}\Big|_{0}\delta g^{kl}(\x_2)+\half\int \d^3\x_2\d^3\x_3
\frac{\delta^2\<T_{pj}(\x_1)\>}{\delta g^{kl}(\x_2)\delta g^{mn}(\x_3)}\Big|_{0}\delta g^{kl}(\x_2)\delta g^{mn}(\x_3),
\]
the zero subscripts indicating setting the sources to their background value ({\it i.e.}, setting $g_{ij}=\delta_{ij}$).
Evaluating this carefully, we have
\begin{align}
\label{big_result}
\delta\<T^i_j(\x_1)\>_s
&= -\half\int\d^3\x_2\<T_{pj}(\x_1)T_{kl}(\x_2)\>\delta^{ip}\delta g^{kl}(\x_2) \nn \\[1ex]
& +\frac{1}{8}\int\d^3\x_2\d^3\x_3 \Big[ \<T_{pj}(\x_1)T_{kl}(\x_2)T_{mn}(\x_3)\> \nn\\[1ex]
& \qquad \qquad -4\delta(\x_1-\x_3)\<T_{mj}(\x_1)T_{kl}(\x_2)\>\delta_{pn}
+\delta(\x_2-\x_3)\<T_{pj}(\x_1)T_{kl}(\x_2)\>\delta_{mn} \nn \\[1ex]
& \qquad \qquad -2\<T_{pj}(\x_1)\Upsilon_{klmn}(\x_2,\x_3)\>-4\<\Upsilon_{pjmn}(\x_1,\x_3)T_{kl}(\x_2)\>
\Big]\delta^{ip}\delta g^{kl}(\x_2)\delta g^{mn}(\x_3), 
\end{align}
where symmetrisation of the quadratic terms under exchange of $\x_2$ and $\x_3$ is understood
and we have dropped ultralocal contact terms but retained semi-local ones.
In addition, we have defined the operator
\[
\label{Upsilon_def}
 \Upsilon_{ijkl}(\x_1,\x_2) = \frac{\delta T_{ij}(\x_1)}{\delta g^{kl}(\x_2)}\Big|_{0} =
2\frac{\delta^2 S}{\delta g^{ij}(\x_1)\delta g^{kl}(\x_2)}\Big|_{0}+\half T_{ij}(\x_1)\delta_{kl}\delta(\x_1-\x_2).
\]

Note that $\psi$ couples to the trace of the stress-energy tensor while
$\gamma_{ij}$ couples to the transverse traceless part.  Hence,
in our holographic calculations to follow, we will only need to turn
on these terms in \eqref{big_result}:
\[
 \delta g_{ij}(\x) = -2\psi\delta_{ij}+\g_{ij}  \quad \Rightarrow \quad \delta g^{ij}(\x) =
 2\psi\delta_{ij}-\g_{ij}+4\psi^2\delta_{ij}-4\psi\g_{ij}+\g_{ik}\g_{kj}.
\]
Transforming to momentum space
and collecting together the coefficients of the various quadratic terms
that appear, we find the variation of the trace of the 1-point function is 
\begin{align}
\label{1ptTrexp}
 \delta\<T^i_i(\vbq_1)\>_s &= - \<\!\<T(\bq_1)T(-\bq_1)\>\!\>\psi(\vbq_1) + \int[[\d\bq_2\d\bq_3]]\big[\ldots\big] \psi(-\vbq_2)\psi(-\vbq_3)\nn\\
& +\int[[\d\bq_2\d\bq_3]]\Big[
-\<\!\<T(\bq_1)T(\bq_2)T^{(s_3)}(\bq_3)\>\!\>
+\Theta_1^{(s_3)}(\bq_i)\<\!\<T(\bq_1)T(-\bq_1)\>\!\> \nn\\[1ex]
& \qquad +\frac{1}{2}\Theta_2^{(s_3)}(\bq_i)\<\!\<T(\bq_2)T(-\bq_2)\>\!\>
+2\<\!\<\Upsilon(\bq_1,\bq_2)T^{(s_3)}(\bq_3)\>\!\> \nn\\[1ex]
&\qquad +2\<\!\<T(\bq_1)\Upsilon^{(s_3)}(\bq_2,\bq_3)\>\!\>
+2\<\!\<T(\bq_2)\Upsilon^{(s_3)}(\bq_1,\bq_3)\>\!\>
\Big]\psi(-\vbq_2)\g^{(s_3)}(-\vbq_3) \nn\\[2ex]
& +\int[[\d\bq_2\d\bq_3]]\Big[
\half\<\!\<T(\bq_1)T^{(s_2)}(\bq_2)T^{(s_3)}(\bq_3)\>\!\>
-\frac{1}{4}\big(A(\bq_2)+A(\bq_3)\big)\theta^{(s_2s_3)}(\bq_i) \nn\\[1ex]
& \qquad-\frac{1}{4}\<\!\<T(\bq_1)T(-\bq_1)\>\!\>\Theta^{(s_2s_3)}(\bq_i)
-\<\!\<T(\bq_1)\Upsilon^{(s_2s_3)}(\bq_2,\bq_3)\>\!\> \nn\\[1ex]
&\qquad -\<\!\<T^{(s_3)}(\bq_3)\Upsilon^{(s_2)}(\bq_1,\bq_2)\>\!\>
 -\<\!\<T^{(s_2)}(\bq_2)\Upsilon^{(s_3)}(\bq_1,\bq_3)\>\!\>
\Big]\g^{(s_2)}(-\vbq_2)\g^{(s_3)}(-\vbq_3),
\end{align}
where we have omitted the coefficient of the $\psi\psi$ term as we
will not need it in the following.  (See instead
\cite{McFadden:2010vh}).  A precise definition of the various
quantities appearing in this expression is given in Appendix
\ref{App_Conventions}. Had we considered only correlators at 
separated points, the r.h.s.~of (\ref{1ptTrexp}) would only 
contain the terms with $\<\!\<T(\bq_1)T(-\bq_1)\>\!\>$, 
$\<\!\<T(\bq_1)T(\bq_2)T^{(s_3)}(\bq_3)\>\!\>$ and 
$\<\!\<T(\bq_1)T^{(s_2)}(\bq_2)T^{(s_3)}(\bq_3)\>\!\>$
(so indeed $\psi$ and $ \gamma^{(s)}$ insert $T$ and $T^{(s)}$ respectively).
As mentioned earlier, however, semi-local terms are important and so retain 
these terms as well. 

We will also need the corresponding result for the
variation of the transverse traceless part of the 1-point function,
which reads
\begin{align}
\label{1ptTTexp}
 &\delta\<T^{(s_1)}(\vbq_1)\>_s \equiv \half\,\ep^{(s_1)}_{ij}(-\vbq_1)\delta\<T^i_j(\vbq_1)\> \nn\\[1ex]
&\qquad = \half A(\bq_1)\g^{(s_1)}(\vbq_1) +\int[[\d\bq_2\d\bq_3]]\Big[
\half\<\!\<T^{(s_1)}(\bq_1)T(\bq_2)T(\bq_3)\>\!\>
-\frac{5}{8}\Theta_2^{(s_1)}(\bq_i)\<\!\<T(\bq_2)T(-\bq_2)\>\!\> \nn\\[1ex]
&\qquad\qquad\qquad\quad -\frac{5}{8}\Theta_3^{(s_1)}(\bq_i)\<\!\<T(\bq_3)T(-\bq_3)\>\!\>
 -\<\!\<T^{(s_1)}(\bq_1)\Upsilon(\bq_2,\bq_3)\>\!\>
 -\<\!\<T(\bq_2)\Upsilon^{(s_1)}(\bq_3,\bq_1)\>\!\> \nn\\[1ex]
&\qquad\qquad\qquad\quad -\<\!\<T(\bq_3)\Upsilon^{(s_1)}(\bq_2,\bq_1)\>\!\>
\Big]\psi(-\vbq_2)\psi(-\vbq_3)\nn\\[1ex]
&\qquad +\int[[\d\bq_2\bq_3]]\Big[
-\<\!\<T^{(s_1)}(\bq_1)T^{(s_2)}(\bq_2)T(\bq_3)\>\!\>
+\big(A(\bq_1)+\frac{5}{4} A(\bq_2)\big)\theta^{(s_1s_2)}(\bq_i) \nn\\[1ex]
&\qquad\qquad\qquad\qquad  + 2\<\!\<T^{(s_1)}(\bq_1)\Upsilon^{(s_2)}(\bq_3,\bq_2)\>\!\>
+2\<\!\<T^{(s_2)}(\bq_2)\Upsilon^{(s_1)}(\bq_3,\bq_1)\>\!\> \nn\\[1ex]
&\qquad\qquad\qquad\qquad +2\<\!\<\Upsilon^{(s_1s_2)}(\bq_1,\bq_2)T(\bq_3)\>\!\>
\Big]\g^{(s_2)}(-\vbq_2)\psi(-\vbq_3) \nn\\[1ex]
&\qquad +\int[[\d\bq_2\bq_3]]\Big[
\half\<\!\<T^{(s_1)}(\bq_1)T^{(s_2)}(\bq_2)T^{(s_3)}(\bq_3)\>\!\>
-\frac{1}{8}\big(2A(\bq_1)+A(\bq_2)+A(\bq_3)\big)\Theta^{(s_1s_2s_3)}(\bq_i) \nn\\[1ex]
&\qquad\qquad\qquad\qquad -\Big(\<\!\<T^{(s_1)}(\bq_1)\Upsilon^{(s_2s_3)}(\bq_2,\bq_3)\>\!\>+\mathrm{cyclic\,perms.}\Big)
\Big]\g^{(s_2)}(-\vbq_2)\g^{(s_3)}(-\vbq_3).
\end{align}

In the next section, we will use these formulae to read off components of the stress-energy tensor 3-point function from the asymptotic behaviour of the bulk domain-wall perturbations.

\subsection{Holographic analysis}
\label{Hol_section}

We now discuss the calculation of holographic 3-point functions for
domain-wall spacetimes that are asymptotically AdS, deferring the
discussion of asymptotically power-law domain-walls to Section
\ref{powerlaw_section}.

Working in synchronous (Fefferman-Graham) gauge where $N_i=0$ and
$N=1$, for asymptotically AdS domain-walls we have \cite{Papadimitriou:2004ap}
\[   \label{holT}
\<T^i_j\>_s = \Big[\frac{-2}{\sqrt{g}}\Pi^i_j\Big]_{(3)} = \bK^{-2}\big[K\delta^i_j-K^i_j\big]_{(3)}.
\]
The subscript here indicates that one should select the piece with the 
indicated weight under scale transformations. More precisely, asymptotically 
AdS spacetimes possess a dilatation operator 
(realised asymptotically as the radial derivative 
in Fefferman-Graham coordinates) and one may decompose all covariant 
quantities 
into a sum of terms each having a definite scaling dimension.  
In particular, one can do this for the radial canonical momentum, 
and (\ref{holT}) then instructs 
us to pick the piece with dilatation weight 3. The terms with lower 
weight diverge and keeping only the terms with weight 3 is equivalent to 
holographic renormalization \cite{deHaro:2000xn}.

Our task now is to compute (\ref{holT}) to quadratic order in the sources.
Upon varying, we obtain
\begin{align}
\label{1ptfnTr}
 \delta\<T^i_i(\vbq_1)\>_s &= 2\bK^{-2}[\delta K(\vbq_1)]_{(3)} = \bK^{-2}\Big[\dot{h}(\vbq_1)-\int[[\d \bq_2\d \bq_3]]h_{ij}(-\vbq_2)\dot{h}_{ij}(-\vbq_3)\Big]_{(3)}, \\
\label{1ptfnTT}
 \delta\<T^{(s)}(\vbq_1)\>_s &= -\half\bK^{-2} \ep^{(s)}_{ij}(-\vbq_1)[\delta K^i_j(\vbq_1)]_{(3)} \nn \\
& =\bK^{-2}\Big[-\half\dot{\g}^{(s)}(\vbq_1)+\frac{1}{4}\int[[\d \bq_2\d \bq_3]]\ep^{(s)}_{ij}(-\vbq_1)h_{ik}(-\vbq_2)\dot{h}_{kj}(-\vbq_3)\Big]_{(3)},
\end{align}
where $h=h_{ii}$.
We now wish to expand $\delta\<T^i_i\>_s$ and $\delta\<T^{(s)}\>_s $ 
to quadratic order in $\psi$ and $\g^{(s')}$. This means that we must 
express $\dot{h}_{ij}$ in terms of $\psi$ and $\g^{(s')}$, which we will accomplish
using Hamilton's equations and the definition of the response 
functions. The main complication in performing this step is that the 
system is constrained, and one has to use the constraints
in expressing $\dot{h}_{ij}$ in terms of $\psi$ and $\g^{(s')}$.
In contrast, for a scalar field $\Psi$ in a fixed FRW background, the 
steps that we 
are about to describe are trivial: to linear order
$\pi \sim a^3 \dot{\Psi} \sim \Omega_\Psi \Psi$, where $\Omega_\Psi$
is the response function, and therefore one immediately finds
$\dot{\Psi}$ in terms of $\Psi$.

Let us start by reviewing the computation at linear order.
To this end, we note first that, at linear order, the Hamiltonian and momentum constraints (given in Appendix \ref{App_constraints}) read
\[ \label{const_linear}
  \dot{\psi} = (\ldots)\delta\vphi, \qquad \dot{h} = -\frac{2\bq^2}{a^2H}\psi+\frac{\dot{\vphi}}{H}\delta\dot{\vphi}+(\ldots)\delta\vphi,
\qquad \dot{\omega}_i = 0.
\]
We therefore obtain
 \[
 \dot{h}_{ij}=\frac{\bq_i\bq_j}{\bq^2}\dot{h} +\frac{4}{a^3}\Ei(\bq)\g_{ij}+ (\ldots)\delta\vphi.
 \]
Now, on the one hand, we have
\[ \label{zdot_response}
 \dot{\z}=\frac{1}{2a^3\ep}\Pi=\frac{1}{2a^3\ep}\bOmega_{[2]}(\bq)\z = -\frac{1}{2a^3\ep}\bOmega_{[2]}(\bq)\psi + (\ldots)\delta\vphi,
 \]
and on the other hand,
\[ \label{zdot_linear}
 \dot{\z} = (-\psi-\frac{H}{\dot{\vphi}}\delta\vphi)\dot{} =-\frac{H}{\dot{\vphi}}\delta\dot{\vphi}+(\ldots)\delta\vphi.
 \]
Thus, at linear order,
\[
\label{rules}
 \delta\dot{\vphi} = \frac{H}{a^3\dot{\vphi}}\bOi(\bq)\psi+(\ldots)\delta\vphi, \qquad \dot{h}_{ij}= 
\frac{\bq_i\bq_j}{\bq^2}\(\frac{\bOmega_{[2]}(\bq)}{a^3}-\frac{2\bq^2}{a^2H}\)\psi+\frac{4}{a^3}\Ei(\bq)\g_{ij}+(\ldots)\delta\vphi.
\]
This is all that we need in order to derive the 2-point function (as we will do below). Moreover,
in the calculations to follow, we will use these results to replace all $\delta\dot{\vphi}$ and $\dot{h}_{ij}$ terms appearing in quadratic combinations.

Let us now prepare to do the computations at quadratic order required
for the evaluation of 3-point functions.  The calculation for the
3-point function for the trace of the stress-energy tensor was performed in
\cite{McFadden:2010vh}.  Our principal goal here is then to evaluate
the remaining correlators
$\<\!\<T(\bq_1)T(\bq_2)T^{(s_3)}(\bq_3)\>\!\>$,
$\<\!\<T(\bq_1)T^{(s_2)}(\bq_2)T^{(s_3)}(\bq_3)\>\!\>$ and
$\<\!\<T^{(s_1)}(\bq_1)T^{(s_2)}(\bq_2)T^{(s_3)}(\bq_3)\>\!\>$.  A
useful feature of the first two of these correlators is that they may
both be computed in two different ways: for example, to compute
$\<\!\<T(\bq_1)T(\bq_2)T^{(s_3)}(\bq_3)\>\!\>$, one may either expand
$\delta\<T^i_i(\vbq_1)\>_s$ to quadratic order in
$\psi(-\vbq_2)\g^{(s_3)}(-\vbq_3)$, or one may expand
$\delta\<T^{(s_3)}(\vbq_3)\>_s$ to quadratic order in
$\psi(-\vbq_1)\psi(-\vbq_2)$ (since $\psi$ and $\gamma^{(s)}$ couple to
$T$ and $T^{(s)}$ respectively).  Clearly both these approaches should
yield the same outcome, providing a useful cross-check on our
calculations.

In the following, we will focus on the calculation of the correlator $\<\!\<T(\bq_1)T(\bq_2)T^{(s_3)}(\bq_3)\>\!\>$ via the first of these methods.
As the analysis for the second method as well as that for the remaining 3-point correlators is of a broadly similar nature
we will simply summarise the appropriate results at the end of the present section.
For full details, we refer the reader to Appendix \ref{App_hol}.

Let us start by examining the full Hamiltonian constraint at quadratic order.  From \eqref{Ham_constr},
in position space we have
\[
 (\dot{h}-h_{ij}\dot{h}_{ij}) = \frac{1}{2H}(R_{(1)}+R_{(2)})+\frac{\dot{\vphi}}{H}\delta\dot{\vphi}
-\frac{1}{8H}\dot{h}^2+\frac{1}{8H}\dot{h}_{ij}\dot{h}_{ij}+\frac{1}{2H}\delta\dot{\vphi}^2 +(\ldots)\delta\vphi+ (\ldots)\delta\vphi^2.
\]
The spatial curvature terms $R_{(1)}$ and $R_{(2)}$ are simply local functions of $\psi$, however, (for example,
$R_{(1)}=4a^{-2}\D^2\psi$) and so holographically these terms contribute only ultralocal contact terms to $\delta\<T^i_i(x)\>_s$.
We may therefore discard these terms immediately.  The remaining quadratic terms may then be replaced using \eqref{rules}.  Up to ultralocal contact terms, in momentum space this gives
\begin{align}
\label{eq1}
(\dot{h}-h_{ij}\dot{h}_{ij})(\vbq_1) &= \frac{\dot{\vphi}}{H}\delta\dot{\vphi}(\vbq_1)-\int[[\d \bq_2\d \bq_3]]
\Theta_2^{(s_3)}(\bq_i) \Big(\frac{\bOi(\bq_2)}{a^6 H}-\frac{2\bq_2^2}{a^5 H^2}\Big)\bEi(\bq_3)\psi(-\vbq_2)\g^{(s_3)}(-\vbq_3)\nn \\ &\qquad\qquad\qquad\qquad+\ldots
\end{align}
where we have made use of \eqref{Theta_def} and retained only quadratic terms of the form $\psi(-\vbq_2)\g^{(s_3)}(-\vbq_3)$.

We may now eliminate the $\delta\dot{\vphi}$ as follows.
Firstly, from the gauge-invariant definition \eqref{zeta_gi} of $\z$, in synchronous gauge we have
\begin{align}
\z  &= -\psi  + \frac{1}{2}\pi_{ij}\big(\psi\g_{ij}\big) + \ldots,  \nn\\[1ex]
\dot{\z} &= -\dot{\psi}-\frac{H}{\dot{\vphi}}\delta\dot{\vphi}-2\psi\dot{\psi}
+\frac{H}{\dot{\vphi}^2}\delta\dot{\vphi}\delta\dot{\vphi} \nn \\[1ex]
&\quad +\frac{1}{4}\pi_{ij}\Big[-\frac{\delta\dot{\vphi}}{\dot{\vphi}}\dot{h}_{ij}
-2(\dot{\chi}_{,ki}+\dot{\omega}_{k,i})(-2\psi\delta_{jk}+\g_{jk})-(\dot{\chi}_{,k}+\dot{\omega}_k)(-2\psi_{,k}\delta_{ij}+\g_{ij,k})\nn\\[1ex]
&\qquad\qquad +2\dot{\psi}\g_{ij}+2\psi\dot{\g}_{ij}-\g_{ik}\dot{\g}_{kj}\Big] + \ldots,
\end{align}
where we have omitted terms that vanish when the sources are restricted to $h_{ij}=-2\psi\delta_{ij}+\g_{ij}$, $\delta\vphi=0$.
Upon replacing time-derivatives of perturbations in the quadratic terms using \eqref{rules}, we then find
\begin{align}
\label{z_evaluated}
\z(\vbq_1) &= -\psi(\vbq_1)+\frac{1}{2}\int[[\d\bq_2\d\bq_3]]\Theta_1^{(s_3)}(\bq_i)\psi(-\vbq_2)\g^{(s_3)}(-\vbq_3)+\ldots, \\[1ex]
\label{eq2}
\dot{\z}(\vbq_1) &= -\dot{\psi}(\vbq_1)-\frac{H}{\dot{\vphi}}\delta\dot{\vphi}(\vbq_1)-\int[[\d\bq_2\d\bq_3]]
\Theta_1^{(s_3)}(\bq_i)\,\Big[\frac{1}{2a^6\ep H}\bOi(\bq_2)\bEi(\bq_3) -\frac{2}{a^3}\bEi(\bq_3)
\nn \\
&\qquad\qquad\qquad +\frac{(\bq_2^2-\bq_1^2-3\bq_3^2)}{16 \bq_2^2}\Big(\frac{\bOi(\bq_2)}{a^3}-\frac{2\bq_2^2}{a^2 H}\Big) \Big]\psi(-\vbq_2)\g^{(s_3)}(-\vbq_3) + \ldots
\end{align}

On the other hand, from \eqref{zdot} combined with \eqref{z_evaluated}, we have
\begin{align}
\label{eq3}
 \dot{\z}(\q_1) &= -\frac{1}{2a^3\ep}\Oi(\bq_1)\psi(\vbq_1)-\int[[\d \bq_2\d \bq_3]]\Big[\frac{1}{2a^3\ep}\Oii^{(s_3)}(\bq_i)
- \frac{1}{4a^3\ep}\Theta_1^{(s_3)}(\bq_i)\bOi(\bq_1)+ \mC^{(s_3)}_{213}  \nn\\[1ex]
&\qquad\qquad +\mD^{(s_3)}_{213} \bEi(\bq_3) +2\mE^{(s_3)}_{123}\bOi(\bq_2)+2\mF^{(s_3)}_{123}\bOi(\bq_2)\bEi(\bq_3)\Big]\psi(-\vbq_2)\bg^{(s_3)}(-\vbq_3) +\ldots,
\end{align}

Finally, to eliminate $\dot{\psi}$, we use the momentum constraint equation.  At quadratic order, this reads
\begin{align}
\dot{\psi} &= -\frac{1}{4}(-2\psi\delta_{ij}+\g_{ij})\dot{h}_{ij} \nn\\
&\qquad +\D^{-2}\D_i\Big[
\frac{1}{4}\D_{j}\big((-2\psi\delta_{jk}+\g_{jk})\dot{h}_{ki}\big)+\frac{1}{8}\dot{h}_{jk}(-2\psi_{,i}\delta_{jk}+\g_{jk,i})-\frac{1}{8}\dot{h}_{ij}(-6\psi_{,j})\Big] +\ldots,
\end{align}
where again we omit terms that vanish when the sources are set to $h_{ij}=-2\psi\delta_{ij}+\g_{ij}$ and $\delta\vphi=0$.
Using \eqref{rules} for the quadratic terms, we then find
\begin{align}
\label{eq4}
\dot{\psi}(\vbq_1) &= -\int[[\d\bq_2\d\bq_3]]\Theta_1^{(s_3)}(\bq_i)\,\Big[ \frac{1}{a^3}\bEi(\bq_3)
+\frac{(\bq_2^2-\bq_1^2-\bq_3^2)}{16\bq_2^2}\Big(\frac{\bOi(\bq_2)}{a^3}-\frac{2\bq_2^2}{a^2H}\Big)\Big]\psi(-\vbq_2)\g^{(s_3)}(-\vbq_3) \nn\\ &\qquad\qquad + \ldots
\end{align}

Putting together \eqref{eq1}, \eqref{eq2}, \eqref{eq3} and \eqref{eq4}, we find (after a number of cancellations)
\begin{align}
(\dot{h}-h_{ij}\dot{h}_{ij})(\vbq_1) &= \frac{1}{a^3}\bOi(\bq_1)\psi(\vbq_1)+\int[[\d\bq_2\d\bq_3]]\Big[\frac{1}{a^3}\bOii^{(s_3)}(\bq_i)-\frac{1}{2a^3}\Theta_1^{(s_3)}(\bq_i)\bOi(\bq_1) \nn \\
&\qquad\qquad\qquad -\frac{2}{a^2H}\theta^{(s_3)}(\bq_i)\Big]\psi(-\vbq_2)\g^{(s_3)}(-\vbq_3)+\ldots
\end{align}
This last term, however, is an ultralocal contact term (note that it is finite upon taking any given $\bq_i$ to zero) and so may be dropped with impunity.
Therefore we find
\begin{align}
\delta\<T^i_i(\vbq_1)\>_s &= \bK^{-2}\bar{\Omega}_{[2](0)}(\bq_1)\psi_{(0)}(\vbq_1) \nn \\[1ex] &+\int[[\d\bq_2\d\bq_3]]\Big[\bK^{-2}\bar{\Omega}_{[3](0)}^{(s_3)}(\bq_i)
-\frac{1}{2}\Theta_1^{(s_3)}(\bq_i)\bK^{-2}\bar{\Omega}_{[2](0)}(\bq_1)\Big]\psi_{(0)}(-\vbq_2)\g_{(0)}^{(s_3)}(-\vbq_3)+\ldots
\end{align}

Comparing with \eqref{1ptTrexp}, we then see that
\begin{align}
\label{Omega2_result}
 -\bK^{-2}\bar{\Omega}_{[2](0)}(\bq) &= \<\!\<T(\bq)T(-\bq)\>\!\> \\
\label{cross_check1}
-\bK^{-2}\bar{\Omega}_{[3](0)}^{(s_3)}(\bq_1,\bq_2,\bq_3) &= \<\!\<T(\bq_1)T(\bq_2)T^{(s_3)}(\bq_3)\>\!\>
-\frac{1}{2}\Theta_1^{(s_3)}(\bq_i)\<\!\<T(\bq_1)T(-\bq_1)\>\!\> \nn\\
&\quad  -\frac{1}{2}\Theta_2^{(s_3)}(\bq_i)\<\!\<T(\bq_2)T(-\bq_2)\>\!\>
 -2\<\!\<\Upsilon(\bq_1,\bq_2)T^{(s_3)}(\bq_3)\>\!\> \nn\\[1ex]
&\quad -2\<\!\<T(\bq_1)\Upsilon^{(s_3)}(\bq_2,\bq_3)\>\!\>
-2\<\!\<T(\bq_2)\Upsilon^{(s_3)}(\bq_1,\bq_3)\>\!\>.
\end{align}

The analysis for the remaining response functions (as well as the cross-check calculation for this last result) may be found in Appendix \ref{App_hol}.
Here, we merely present the final results of these calculations, which are
\newpage
\begin{align}
\label{cross_check3}
-\bK^{-2}\bar{\Omega}_{[3](0)}^{(s_2s_3)}(\bq_1,\bq_2,\bq_3) &=
 \half\<\!\<T(\bq_1)T^{(s_2)}(\bq_2)T^{(s_3)}(\bq_3)\>\!\>
-\frac{1}{4}\big(A(\bq_2)+A(\bq_3)\big)\theta^{(s_2s_3)}(\bq_i) \nn\\[1ex]
&\quad -\frac{1}{8}\<\!\<T(\bq_1)T(-\bq_1)\>\!\>\Theta^{(s_2s_3)}(\bq_i)
-\<\!\<T(\bq_1)\Upsilon^{(s_2s_3)}(\bq_2,\bq_3)\>\!\> \nn\\[1ex]
&\quad
-\<\!\<T^{(s_2)}(\bq_2)\Upsilon^{(s_3)}(\bq_1,\bq_3)\>\!\>
-\<\!\<T^{(s_3)}(\bq_3)\Upsilon^{(s_2)}(\bq_1,\bq_2)\>\!\>,
\end{align}
as well as
\begin{align}
\label{E2_result}
 -4\bK^{-2}\bar{E}_{[2](0)}(\bq) &= A(\bq), \\
\label{Esss_result}
-2\bK^{-2}\bar{E}_{[3](0)}^{(s_1s_2s_3)}(\bq_i) &= \half\<\!\<T^{(s_1)}(\bq_1)T^{(s_2)}(\bq_2)T^{(s_3)}(\bq_3)\>\!\>
-\frac{1}{8}\Theta^{(s_1s_2s_3)}(\bq_i)\sum_i A(\bq_i) \nn\\
&\quad -\Big(\<\!\<T^{(s_1)}(\bq_1)\Upsilon^{(s_2s_3)}(\bq_2,bq_3)\>\!\> + \mathrm{2\,\,cyclic\,perms.}\Big).
\end{align}
In these expressions $A(\bq)$ refers to the transverse traceless piece of the stress-energy tensor 2-point function as defined in \eqref{ABdef}.

In summary, the main results of this section are \eqref{Omega2_result}, \eqref{cross_check1}, \eqref{cross_check3}, \eqref{E2_result}.
These results allow us to read off the dual 3-point correlation functions from the asymptotic behaviour of the bulk response functions.

\subsection{Asymptotically power-law domain-walls}
\label{powerlaw_section}

As noted in the Introduction, there are two classes of domain-wall spacetime that currently have a well-understood holographic description:
the first class consists of domain-walls that are asymptotically AdS, for which the holographic analysis is discussed above, while the second class consists of domain-walls that asymptote to non-conformal brane backgrounds.
This latter class of domain-wall solutions correspond to cosmologies that have asymptotic power-law scaling at late times.
For a detailed description of the relevant background geometry and holographic analysis we refer the reader to \cite{Kanitscheider:2008kd,McFadden:2010na, McFadden:2010vh}.
The holographic analysis in particular is very closely to related to that for the asymptotically AdS case.
In fact, in Section 4.2.2 of \cite{McFadden:2010vh}, we showed that the holographic formula giving the 3-point function for the trace of the stress-energy tensor for asymptotically AdS domain-walls also holds in the case of asymptotically power-law domain-walls.
Here, it suffices to note that exactly the same arguments apply in the present case, and our results above expressing the stress-energy tensor 3-point function in terms of the bulk response functions are equally valid for both asymptotically AdS and asymptotically power-law domain-walls.
A brief summary of the arguments of Section 4.2.2 of \cite{McFadden:2010vh} is given below, where we note a few additional points of relevance.

In the asymptotically power-law case, the 1-point function in the presence of sources is given by the canonical momentum in the dual frame \cite{Kanitscheider:2008kd},
\[
\label{DF1ptfn}
\<T^i_j(x)\>_s = \Big[\frac{-2}{\sqrt{\tilde{g}}}\,\tilde{\Pi}^i_j\Big]_{(3)} = \Big[\bK^{-2}e^{\lambda\Phi}\big((\tilde{K}+\lambda\Phi_{,r})\delta^i_j-\tilde{K}^i_j\big)\Big]_{(3)},
\]
where all quantities are defined in Section 4 of \cite{McFadden:2010vh}.
Expanding this 1-point function in the dual frame fluctuations $\tilde{\psi}$ and $\tilde{\g}$ is equivalent to expanding in powers of the Einstein frame fluctuations $\psi$ and $\g$, since the respective coefficients of $\tilde{\psi}$, $\tilde{\g}$, $\tilde{\psi}\tilde{\psi}$, $\tilde{\g}\tilde{\g}$ and $\tilde{\psi}\tilde{\g}$ in the dual frame are equal to the coefficients of $\psi$, $\g$, $\psi\psi$, $\g\g$ and $\psi\g$ in the Einstein frame (see (4.33) of \cite{McFadden:2010vh}).
Expanding \eqref{DF1ptfn} and converting the dual frame perturbations into their Einstein frame equivalents, therefore, we find
\begin{align}
\label{dftrace}
 \delta\<T^i_i(\vbq_1)\>_s &=  \Big[\bK^{-2} e^{3\lambda\vphi/2}\Big(\dot{h}(\vbq_1)-\int[[\d \bq_2\d \bq_3]]h_{ij}(-\vbq_2)\dot{h}_{ij}(-\vbq_3)+\ldots\Big)\Big]_{(3)}, \\[1ex]
\label{dfTT}
 \delta\<T^{(s)}(\vbq_1)\>_s &= \Big[\bK^{-2}e^{3\lambda \vphi/2}\Big({-}\half\dot{\g}^{(s)}(\vbq_1)+\frac{1}{4}\int[[\d \bq_2\d \bq_3]]\ep^{(s)}_{ij}(-\vbq_1)h_{ik}(-\vbq_2)\dot{h}_{kj}(-\vbq_3)+\ldots\Big)\Big]_{(3)},
\end{align}
where we need retain only terms contributing to the expansion in $\psi$ and $\g$.
In particular, these expressions differ from their asymptotically AdS counterparts \eqref{1ptfnTr} and \eqref{1ptfnTT} only by an overall factor of
$e^{3\lambda\vphi/2}$.
We may therefore make use of our previous results for the asymptotically AdS case, noting that the effect of this overall factor is simply to convert the factors of $a^{-3}$ appearing in our previous expressions to factors of $\tilde{a}^{-3}$, where $\tilde{a}$ is the dual frame scale factor.  (In this analysis, it is also important that the gauge-invariant fluctuations $\z$ and $\bg_{ij}$ defined in \eqref{zeta_gi} and \eqref{gb_gi} are independent of the lapse perturbation $\dN$, as discussed in \cite{McFadden:2010vh}).
Consequently, at the end of our manipulations, when we extract the piece with dilatation weight three in the dual frame, we obtain exactly the same result as in the asymptotically AdS case earlier, when we extracted the piece with dilatation weight three in the Einstein frame.  This is because $a^{-3}$ has dilatation weight three in the Einstein frame, while $\tilde{a}^{-3}$ has dilatation weight three in the dual frame.

\section{Cosmological 3-point correlators from holography} 
\label{summary}

In Section \ref{cosmo_calcs}, we saw that the cosmological 2- and 3-point functions are related to the cosmological response functions, while in the previous section, we saw that the domain-wall response functions are related to 2- and 3-point functions of the dual QFT.  
We will now combine these results to obtain the main result of this paper: 
a complete set of holographic formulae for all cosmological 2- and 3-point functions in terms of 2- and 3-point functions of the dual QFT.

First, combining the cosmological 2-point functions \eqref{powerspecresults} evaluated at late times with our holographic results 
\eqref{Omega2_result} and \eqref{E2_result}, we recover the relations \cite{McFadden:2009fg, McFadden:2010na}
\[
\label{2ptformulae}
 \<\!\<\z(q)\z(-q)\>\!\> = \frac{-1}{8\Im[B(\bq)]}, \qquad \<\!\<\bg^{(s)}(q)\bg^{(s')}(-q)\>\!\> = \frac{-\delta^{ss'}}{\Im[A(\bq)]},
\]
where $A(\bq)$ and $B(\bq)$ are respectively the transverse traceless and trace pieces of the stress-energy tensor 2-point function as defined in \eqref{ABdef}.

Next, combining the results \eqref{result1}, \eqref{result2} and \eqref{result3} for the cosmological 3-point functions (also evaluated at late times) with the corresponding holographic results \eqref{cross_check1}, \eqref{cross_check3} and \eqref{Esss_result}, together with \eqref{2ptdefs} and \eqref{2ptformulae}, we find 
\begin{align}
\label{holo_zzz}
&\<\!\<\z(q_1)\z(q_2)\z(q_3)\>\!\>  \nn\\[1ex]&\quad
 = -\frac{1}{256}\Big(\prod_i \Im [B(\bq_i)]\Big)^{-1}\times
\Im \Big[\<\!\<T(\bq_1)T(\bq_2)T(\bq_3)\>\!\> + 4\sum_i B(\bq_i) \nn\\[0ex]&\qquad\qquad\qquad\qquad\qquad\qquad\qquad\qquad
-2\Big( \<\!\<T(\bq_1)\Upsilon(\bq_2,\bq_3)\>\!\>+\mathrm{cyclic\,perms.}\Big)\Big], \nn\\[0ex]
\\[-2ex]
\label{holo_zzg}
&\<\!\<\z(q_1)\z(q_2)\bg^{(s_3)}(q_3)\>\!\>  \nn\\[1ex]& \quad
= -\frac{1}{32} \Big(\Im[B(\bq_1)]\Im[B(\bq_2)]\Im[A(\bq_3)]\Big)^{-1} \nn\\[1ex]& \qquad 
\times \Im\Big[\<\!\<T(\bq_1)T(\bq_2)T^{(s_3)}(\bq_3)\>\!\> 
-2\big(\Theta_1^{(s_3)}(\bq_i)B(\bq_1)+\Theta_2^{(s_3)}(\bq_i)B(\bq_2)\big)  \nn\\[1ex]&\qquad\qquad\quad
-2\<\!\<\Upsilon(\bq_1,\bq_2)T^{(s_3)}(\bq_3)\>\!\> 
-2\<\!\<T(\bq_1)\Upsilon^{(s_3)}(\bq_2,\bq_3)\>\!\>  
-2\<\!\<T(\bq_2)\Upsilon^{(s_3)}(\bq_1,\bq_3)\>\!\>
\Big],  \nn\\[1ex]
\\[0ex]
\label{holo_zgg}
& \<\!\<\z(q_1)\bg^{(s_2)}(q_2)\bg^{(s_3)}(q_3)\>\!\> \nn\\[1ex]&\quad
= -\frac{1}{4}\Big(\Im[B(\bq_1)]\Im[A(\bq_2)]\Im[A(\bq_3)]\Big)^{-1} \nn\\[1ex]&\qquad
\times \Im\Big[\<\!\<T(\bq_1)T^{(s_2)}(\bq_2)T^{(s_3)}(\bq_3)\>\!\> 
-\half \big(A(\bq_2)+A(\bq_3)\big)\theta^{(s_2s_3)}(\bq_i) 
-B(\bq_1) \Theta^{(s_2s_3)}(\bq_i)  \nn\\[1ex]&\qquad\quad
-2\<\!\<T(\bq_1)\Upsilon^{(s_2s_3)}(\bq_2,\bq_3)\>\!\> 
-2\<\!\<T^{(s_3)}(\bq_3)\Upsilon^{(s_2)}(\bq_1,\bq_2)\>\!\>
-2\<\!\<T^{(s_2)}(\bq_2)\Upsilon^{(s_3)}(\bq_1,\bq_3)\>\!\> \Big], \nn\\[1ex]
\\[0ex]
\label{holo_ggg}
&\<\!\<\bg^{(s_1)}(q_1)\bg^{(s_2)}(q_2)\bg^{(s_3)}(q_3)\>\!\> \nn\\[1ex]&\quad
= -\Big(\prod_i\Im[A(\bq_i)]\Big)^{-1}   
\times\Im \Big[2\<\!\<T^{(s_1)}(\bq_1)T^{(s_2)}(\bq_2)T^{(s_3)}(\bq_3)\>\!\> 
-\frac{1}{2}\Theta^{(s_1s_2s_3)}(\bq_i)\sum_i A(\bq_i) \nn\\[1ex]&\qquad \qquad\qquad\qquad \qquad\qquad\qquad 
-4\Big(\<\!\<T^{(s_1)}(\bq_1)\Upsilon^{(s_2s_3)}(\bq_2,bq_3)\>\!\> + \mathrm{cyclic\,perms.}\Big)
 \Big]. \nn\\[0ex]
\end{align}
The imaginary part in these formulae is taken after the analytic continuation \eqref{cont} or \eqref{QFT_cont} is made.
Our notation for the various correlators is given in Appendix \ref{App_Conventions}, while the contractions of helicity tensors appearing in these formulae are given in Appendix \ref{App_helicity}.  The operator $\Upsilon$ was defined in \eqref{Upsilon_def}, and its symmetry properties are discussed in Appendix \ref{App_Conventions}.
For completeness, we have added here the formula \eqref{holo_zzz} for the 3-point function of $\z$ as derived in \cite{McFadden:2010vh}.

Note that all quantities appearing on the r.h.s.~of these formulae relate to the dual QFT.  Each r.h.s.~consists of an overall prefactor 
constructed from the 2-point function multiplying a sum of the appropriate 3-point function along with various semi-local terms.
The semi-local terms vanish when all operators are at separate points, but they may be non-zero if two of the operators are coincident.
In the case of \eqref{holo_zzz}, it was shown in \cite{McFadden:2010vh} that these semi-local terms contribute to `local'-type non-Gaussianity. 


\section{Discussion \label{sec:discussion}}

In this paper we studied tree-level in-in cosmological 3-point functions for single scalar universes, 
including both scalar and tensor perturbations.
For cosmologies that are either asymptotically dS or asymptotically power-law at late times,
we showed that these 3-point functions may be re-expressed in terms of
the stress-energy tensor correlation functions of a dual QFT.
These holographic formulae are our main results and are collected in Section \ref{summary}.

Let us first discuss the correlators appearing in these formulae from the perspective of the dual QFT.
Stress-energy tensor correlation functions are defined by 
coupling the QFT to a background metric and then successively functionally
differentiating with respect to the metric, before setting the background metric equal
to the flat metric. Functionally differentiating, say, three times
with respect to the background metric
gives rise, in addition to the 3-point function
of $T_{ij}$, to semi-local\footnote{In semi-local terms two of the
three insertion point are coincident, while in ultralocal terms all
insertion points are coincident.} and ultralocal terms since the
stress-energy tensor itself depends on the background metric.
The ultralocal terms are not important (except when they are related to
anomalies, but there are no relevant anomalies in the case at hand)
because their value can be changed at will by adding a finite
local counterterm. On the other hand, semi-local terms are important.
From a cosmological perspective, they may contribute to non-Gaussianity of the
`local' type \cite{McFadden:2010vh}. 
Our holographic formulae therefore carefully include the contribution
of all such terms.

Note that on the holographic side, the results presented here are
the complete 3-point functions involving the stress-energy tensor.
Correlation functions involving the stress-energy tensor and the
scalar operator dual to the bulk scalar field 
follow from Ward identities \cite{Kanitscheider:2008kd}.

We found it useful to adopt a helicity basis for the tensor perturbations.
When the bulk action is helicity preserving then one only needs to specify
the correlators with zero or one negative helicity graviton. The rest then follow
by permutations and/or a parity transformation. Furthermore,
in all single scalar inflationary models
based on Einstein gravity (with canonical kinetic terms for the scalars),
the ratios of the 3-point functions involving only positive helicity gravitons
to their counterparts with one negative helicity graviton are universal,
and are given by a ratio of momenta that is independent of the potential,
see (\ref{consistency}). Thus, all correlators that involve tensors
are determined from those with only positive helicity gravitons.

The holographic formulae derived here may also be used to extract
predictions for holographic models of inflation in which the very early
universe is in a non-geometric strongly coupled phase. To achieve this,
one needs to compute the relevant QFT correlators in perturbation theory.
A class of models analysed in our previous papers
correspond to universes that at late times are described by a
power-law geometry (where late time 
refers here to the end of the holographic epoch, which is also the beginning of standard
hot big bang cosmology). The dual theory is described by an $SU(N)$
Yang-Mills theory coupled to massless scalars and fermions
with only Yukawa-type and quartic scalar interaction terms, 
and the relevant leading-order computation amounts to a 1-loop computation.
This computation, and the corresponding predictions for the cosmological bispectra,
will be discussed elsewhere \cite{toappear}.

\bigskip
{\it Acknowledgments:} The authors are supported by NWO, the Nederlandse Organisatie voor Wetenschappelijk Onderzoek.


\appendix


\section{Gauge-invariant variables at second order}
\label{App_GT}

In this appendix we derive gauge-invariant definitions of the variables $\zeta$ and $\bg_{ij}$.
Decomposing the metric to second order as $g_{\mu\nu} = g^{(0)}_{\mu\nu}+\delta g_{\mu\nu}$, a generic metric perturbation $\delta g_{\mu\nu}$ transforms under a gauge transformation $\xi^\mu$ as
\[
\label{myGT}
 \delta\check{g}_{\mu\nu} = \delta g_{\mu\nu}+\Lie_\xi g^{(0)}_{\mu\nu}+\Lie_\xi \delta g_{\mu\nu}+
 \frac{1}{2} \Lie^2_\xi g^{(0)}_{\mu\nu}.
\]
Note that upon setting
\[
\label{GTexpansion}
\delta g_{\mu\nu}=\lambda \delta g_{\mu\nu}^{(1)}+\frac{\lambda^2}{2}\, \delta g_{\mu\nu}^{(2)}+O(\lambda^3), \qquad \xi^\mu = \lambda \xi_{(1)}^\mu+\frac{\lambda^2}{2}\, \xi_{(2)}^\mu + O(\lambda^3),
\]
and expanding in powers of $\lambda$, \eqref{myGT} may equivalently be written \cite{Bruni:1996im, Matarrese:1997ay}
\[
\label{BruniGT}
\delta\check{g}_{\mu\nu}^{(1)} = \delta g_{\mu\nu}^{(1)}+\Lie_{\xi_{(1)}}g_{\mu\nu}^{(0)}, \qquad
\delta \check{g}_{\mu\nu}^{(2)} = \delta g_{\mu\nu}^{(2)}+\Lie_{\xi_{(2)}}g_{\mu\nu}^{(0)}
+\Lie^2_{\xi_{(1)}}g_{\mu\nu}^{(0)}+2\Lie_{\xi_{(1)}}\delta g_{\mu\nu}^{(1)}.
\]
The transformed metric perturbations, as defined in \eqref{00metric_perts} and \eqref{metric_perts}, are then
\begin{align}
&\check{\phi} = (1/2)\sigma\delta\check{g}_{00},  \qquad
&\check{\nu}_i = a^{-2}\pi_{ij}\delta\check{g}_{0j},  \nn \\
&\check{\nu} = a^{-2}\D^{-2}\D_i \delta\check{g}_{0i}, \qquad
&\check{\omega}_i = a^{-2}\pi_{ij}\D_k \D^{-2}\delta\check{g}_{jk}, \nn\\
&\check{\psi} = -(1/4)a^{-2}\pi_{ij}\delta\check{g}_{ij}, \qquad
&\check{\g}_{ij} = a^{-2} \Pi_{ijkl}\delta\check{g}_{kl}, \nn\\
&\check{\chi} = (1/2)a^{-2}(\delta_{ij}-(3/2)\pi_{ij})\D^{-2}\delta\check{g}_{ij},
\end{align}
where the transverse and transverse traceless projection operators $\pi_{ij}$ and $\Pi_{ijkl}$ are defined in \eqref{projection_operators}.

These formulae may be evaluated explicitly as required.  In the following, we will need $\check{\chi}$ and $\check{\omega}_i$ to first order, and $\check{\psi}$ and $\check{\g}_{ij}$ to second order.
Writing $\xi^\mu = (\alpha, \delta^{ij}\xi_j)$ (where $\xi_i$ may be further decomposed as $\xi_i = \beta_{,i}+\g_i$, where $\g_i$ is transverse), we find that at first order
\[
 \check{\chi} = \chi+\beta, \qquad \check{\omega}_i = \omega_i+\gamma_i,
\]
while at second order
\begin{align}
\label{psi_transf}
 \check{\psi} &= \psi - H\alpha -\big(\frac{\dot{H}}{2}+H^2\big)\alpha^2-\frac{H}{2}\alpha\dot{\alpha}-\frac{H}{2}\xi_i\alpha_{,i}-\frac{1}{4}\pi_{ij}Y_{ij},  \\
\label{g_transf}
\check{\g}_{ij} &= \g_{ij}+\Pi_{ijkl}Y_{kl}.
\end{align}
Here, the quadratic combination
\[
Y_{ij}= \alpha\dot{h}_{ij}+2H\alpha h_{ij}+\xi_k h_{ij,k}+\frac{2}{a^2} \dN_i\alpha_{,j}+2\xi_{k,i}h_{jk}+\frac{\sigma}{a^2} \alpha_{,i}\alpha_{,j}
+\xi_{k,i}\xi_{k,j}+4H\alpha\xi_{i,j}.
\]
We will also need the transformation of the scalar field perturbation to second order, 
\bea
\label{scalar_transf}
 \delta\check{\vphi} &=& \delta\vphi +\Lie_\xi \vphi +\Lie_\xi \delta \vphi+(1/2) \Lie^2_\xi \vphi \nn \\
&=& \delta\vphi + \alpha\dot{\vphi}+\alpha\delta\dot{\vphi}+\xi_i\delta\vphi_{,i}+(1/2)\ddot{\vphi}\alpha^2+(1/2)\dot{\vphi}\alpha\dot{\alpha}+(1/2)\dot{\vphi}\xi_i\alpha_{,i}.
\eea

To identify the gauge-invariant definitions of $\z$ and $\bg_{ij}$, we consider transforming from a general gauge to the comoving gauge in which
\[
\label{defn_comoving_gauge}
g^{co}_{ij} = a^2 e^{2\z}[e^{\bg}]_{ij} 
=a^2[\delta_{ij}+(2\z\delta_{ij}+\bg_{ij}) +(2\z^2\delta_{ij}+2\z\bg_{ij}+\frac{1}{2}\bg_{ik}\bg_{kj})], \quad \delta\vphi^{co} = 0.
\]
Recalling that $\bg_{ij}$ is transverse traceless, to first order this requires $\alpha = -\delta\vphi/\dot{\vphi}$ and $\xi_i =-(\chi_{,i}+\omega_i)$.
Using these first order quantities, we may then solve \eqref{scalar_transf} to quadratic order, whence 
\[
\label{alpha_second}
 \alpha = -\frac{\delta\vphi}{\dot{\vphi}}+\frac{\delta\vphi\delta\dot{\vphi}}{2\dot{\vphi}^2}+(\chi_{,i}+\omega_i)\frac{\delta\vphi_{,i}}{2\dot{\vphi}}.
\]
Knowing $\alpha$ to second order and $\xi_i$ to first order, we now have sufficient information to identify
$\psi^{co}$ and $\g^{co}_{ij}$ to second order using \eqref{psi_transf} and \eqref{g_transf}.
On the other hand, from \eqref{defn_comoving_gauge}, we have
\[
 \psi^{co} = -\z-\z^2-\frac{1}{4}\pi_{ij} \big( 2\z\bg_{ij}+\frac{1}{2}\bg_{ik}\bg_{kj}\big), \qquad
\g^{co}_{ij} = \bg_{ij}+\Pi_{ijkl}\big( 2\z\bg_{ij}+\frac{1}{2}\bg_{ik}\bg_{kj}\big),
\]
which, upon inverting, yields
\[
 \z = -\psi^{co}-(\psi^{co})^2+\frac{1}{4}\pi_{ij}(2\psi^{co}\g^{co}_{ij}-\frac{1}{2}\g^{co}_{ik}\g^{co}_{kj}), \qquad
\bg_{ij} = \g^{co}_{ij}+\Pi_{ijkl}(2\psi^{co}\g^{co}_{kl}-\frac{1}{2}\g^{co}_{km}\g^{co}_{ml}).
\]
We may thus write down $\z$ and $\bg_{ij}$ to quadratic order; the result is given in \eqref{zeta_gi} and \eqref{gb_gi}.

Finally, let us note that the gauge \eqref{defn_comoving_gauge} is fully fixed: in addition to \eqref{alpha_second}, there is a unique solution for $\xi_i$ at quadratic order which may obtained by evaluating $\check{\chi}$ and $\check{\omega}$ to second order and matching to \eqref{defn_comoving_gauge}.  We have checked this explicitly, along with the gauge-invariance of our final expressions for $\z$ and $\bg_{ij}$.

\section{Cubic interaction terms}
\label{App_int}

Here we give the action for the perturbations to cubic order and list the various coefficients appearing in the interaction Hamiltonian.
The case $\sigma=+1$ corresponds to domain-walls while $\sigma=-1$ corresponds to cosmologies.

The action is given by
\[
S = \int \d^4 x\, (\mathcal{L}^{(2)}+\mathcal{L}^{(3)}), \qquad \mathcal{L}^{(3)} = \mathcal{L}_{\z\z\z}+\mathcal{L}_{\z\z\bg}+\mathcal{L}_{\z\bg\bg}+\mathcal{L}_{\bg\bg\bg},
\]
where
\begin{align}
\K^2\mathcal{L}^{(2)} &= a^3\ep \dot{\z}^2+\sigma a\ep(\D\z)^2 +\frac{a^3}{8}\dot{\bg}_{ij}\dot{\bg}_{ij}+\frac{\sigma a}{8}\bg_{ij,k}\bg_{ij,k},  \nn\\[1ex]
\K^2\mathcal{L}_{\z\z\z} &= - \frac{a^3\ep}{H}\dot{\z}^3+3a^3\ep\z\dot{\z}^2+\sigma a\ep\z(\D\z)^2
-2a^3\z_{,k}\hat{\nu}_{,k}\D^2\hat{\nu}  -\frac{a^3}{2}\big(\frac{\dot{\z}}{H}-3\z\big) \(\hat{\nu}_{,ij}\hat{\nu}_{,ij}-\D^2 \hat{\nu} \D^2\hat{\nu}\), \nn\\[1ex]
\K^2\mathcal{L}_{\z\z\bg} &= \frac{2\sigma a}{H}\bg_{ij}\dot{\z}_{,i}\z_{,j}+\sigma a \bg_{ij}\z_{,i}\z_{,j}-\frac{a^3}{2} \big(3\z-\frac{\dot{\z}}{H}\big)\dot{\bg}_{ij}\hat{\nu}_{ij}+\frac{a^3}{2} \bg_{ij,k}\hat{\nu}_{ij}\hat{\nu}_{,k},  \nn\\[1ex]
\K^2\mathcal{L}_{\z\bg\bg} &= \frac{a^3}{8}\big(3\z-\frac{\dot{\z}}{H}\big)\dot{\bg}_{ij}\dot{\bg}_{ij}+\frac{\sigma a}{8}(\z+\frac{\dot{\z}}{H})\bg_{ij,k}\bg_{ij,k}-\frac{a^3\ep}{4}\dot{\bg}_{ij}\bg_{ij,k}\D^{-2}\dot{\z}_{,k}-\frac{\sigma a}{4H}\dot{\bg}_{ij}\bg_{ij,k}\z_{,k}, \nn\\[1ex]
\K^2\mathcal{L}_{\bg\bg\bg} &= \frac{\sigma a}{8}\bg_{ij,k}\bg_{ij,l}\bg_{kl}-\frac{\sigma a}{4}\bg_{ij,k}\bg_{kl}\bg_{li,j},
\end{align}
where $\hat{\nu} = \ep\, \D^{-2}\dot{\z}+(\sigma/a^2 H)\z$ and $\ep=-\dot{H}/H^2$. (Note however we are not assuming slow-roll).

From the action we may derive the cubic interaction Hamiltonian.
The coefficients associated with $H_{\z\z\bg}$ defined in \eqref{H_zzg} are
\begin{align}
\label{zzg_coeffts}
&\mA^{(s_3)}_{123} = \Big(\frac{q_3^2}{4aH^2}-\sigma a\Big)\,\theta^{(s_3)}(q_i), \qquad
&&\mD^{(s_3)}_{123} = \Big(\frac{3}{a^3 q_2^2}+\frac{\sigma}{a^5\ep H^2}\Big)\,\theta^{(s_3)}(q_i), \nn\\[1ex]
&\mB^{(s_3)}_{123} = -\frac{6\sigma}{a^2 H}\,\theta^{(s_3)}(q_i), \qquad
&&\mE^{(s_3)}_{123} = \frac{q_3^2}{16 a^3 q_1^2 q_2^2}\,\theta^{(s_3)}(q_i), \nn\\[1ex]
&\mC^{(s_3)}_{123} = -\frac{\sigma}{a^2H}\Big(\frac{1}{\ep}+\frac{q_3^2}{4q_2^2}\Big)\,\theta^{(s_3)}(q_i), \qquad
&&\mF^{(s_3)}_{123} = -\frac{1}{4 a^6 \ep H}\Big(\frac{1}{q_1^2}+\frac{1}{q_2^2}\Big)\theta^{(s_3)}(q_i),
\end{align}
where the shorthand $\mA^{(s_3)}_{123}$ should be understood as $\mA^{(s_3)}(q_1,q_2,q_3)$, {\it etc.}.
In these expressions, and in those below, $\theta^{(s_3)}(q_i)$, $\theta^{(s_2 s_3)}(q_i)$ and $\theta^{(s_1 s_2 s_3)}(q_i)$ denote specific
contractions of helicity tensors which are given in Appendix \ref{App_helicity}.
(Note they are equivalent to real functions of the magnitudes $q_i$ and the helicities $s_i$).
The coefficients associated with $H_{\z\bg\bg}$ defined in \eqref{H_zgg} are
\begin{align}
\label{zgg_coeffts}
&\mA^{(s_2 s_3)}_{123} = \frac{\sigma a}{16}(q_1^2-q_2^2-q_3^2)\,\theta^{(s_2s_3)}(q_i), \qquad
&&\mD^{(s_2 s_3)}_{123} =  \frac{\sigma}{32 a^2 \ep H}(q_1^2-q_2^2-q_3^2)\,\theta^{(s_2s_3)}(q_i), \nn \\
&\mB^{(s_2 s_3)}_{123} = \frac{\sigma}{2 a^2 H}(q_1^2+q_2^2-q_3^2)\,\theta^{(s_2s_3)}(q_i), \qquad
&&\mE^{(s_2 s_3)}_{123} =  \frac{1}{4 a^3 q_1^2}(q_3^2-q_1^2-q_2^2)\,\theta^{(s_2s_3)}(q_i),\nn\\
& \mC^{(s_2s_3)}_{123} = -\frac{6}{a^3}\,\theta^{(s_2s_3)}(q_i), \qquad
&& \mF^{(s_2s_3)}_{123} = \frac{1}{a^6\ep H}\,\theta^{(s_2s_3)}(q_i).
\end{align}
Finally, the single coefficient appearing in \eqref{H_ggg} for $H_{\bg\bg\bg}$ is
\[
\label{ggg_coeffts}
\mA^{(s_1s_2s_3)}(q_i) =  \frac{\sigma a}{24}\theta^{(s_1s_2s_3)}(q_i).
\]

\section{Helicity tensors} 
\label{App_helicity}

This appendix summarises our notation for the various contractions of helicity tensors that appear in the main text.
We also give explicit formulae for these contractions in terms of the magnitudes $q_i$ of the momenta and the helicities $s_i$.

The contractions appearing in the cubic interaction Hamiltonian are
\begin{align}
\label{theta_def}
\theta^{(s_3)}(q_i) &= \ep^{(s_3)}_{ij}(-\q_3)q_1^i q_1^j = \ep^{(s_3)}_{ij}(-\q_3)q_2^i q_2^j, \nn\\
 \theta^{(s_2s_3)}(q_i) &= \ep^{(s_2)}_{ij}(-\q_2)\ep^{(s_3)}_{ij}(-\q_3),  \nn\\
\theta^{(s_1s_2s_3)}(q_i) &= \ep^{(s_1)}_{ii'}(-\q_1)\ep^{(s_2)}_{jj'}(-\q_2)\ep^{(s_3)}_{kk'}(-\q_3) t_{ijk} t_{i'j'k'},
\end{align}
where $t_{ijk} = \delta_{ij}q_{1k}+\delta_{jk}q_{2i}+\delta_{ki}q_{3j}$.
In addition, the following contractions arise in the holographic analysis
\begin{align}
\label{Theta_def}
&\Theta_1^{(s_3)}(\bq_i) = \pi_{ij}(\bq_1)\ep_{ij}^{(s_3)}(-\bq_3), \qquad
&& \Theta^{(s_2s_3)}(\bq_i) = \pi_{ij}(\bq_1)\ep_{ik}^{(s_2)}(-\vbq_2)\ep_{kj}^{(s_3)}(-\vbq_3), \nn\\
& \Theta_2^{(s_3)}(\bq_i) = \pi_{ij}(\bq_2)\ep_{ij}^{(s_3)}(-\bq_3), \qquad
&&\Theta^{(s_1s_2s_3)}(\bq_i) = \ep^{(s_1)}_{ij}(-\vbq_1)\ep^{(s_2)}_{jk}(-\vbq_2)\ep^{(s_3)}_{ki}(-\vbq_3).
\end{align}

To evaluate these contractions explicitly, it is useful to introduce an explicit basis of helicity tensors.
The analysis is simplified by the fact that 
all momenta lie in a single plane due to momentum conservation.
Taking this plane to be the $(x,y)$ plane, we have
\[
 \q_1 = q_1\,(1,0,0),\quad \q_2=q_2\,(\cos\theta, \sin\theta,0), \quad \q_3=q_3\,(\cos\phi,\sin\phi,0),
\]
where the magnitudes $q_i\ge 0$, and without loss of generality we may choose $0\le\theta\le\pi$ and $\pi\le\phi\le2\pi$ so that
\begin{align}
 \cos\theta = \frac{(q_3^2-q_1^2-q_2^2)}{2q_1q_2}, \quad \sin\theta = \frac{\lambda}{2q_1q_2}, \quad 
\cos\phi = \frac{(q_2^2-q_1^2-q_3^2)}{2q_1q_3}, \quad \sin\phi = -\frac{\lambda}{2q_1q_3},
\end{align}
where
\[
\label{lambda_def}
\lambda = +\sqrt{2q_1^2q_2^2+2q_2^2q_3^2+2q_3^2q_1^2-q_1^4-q_2^4-q_3^4}.
\]
The required helicity tensors may then be found by rotation in the $(x,y)$ plane:
\vspace{0.3cm}
\begin{align}
& \ep^{(s_1)}(\q_1) = \frac{1}{\sqrt{2}}\left(\begin{array}{ccc} 0 & 0 & 0 \\ 0 & 1 & is_1 \\ 0 & is_1 & -1 \end{array}\right), \qquad
 \ep^{(s_2)}(\q_2) = \frac{1}{\sqrt{2}}\left(\begin{array}{ccc} \sin^2\theta & -\sin\theta\cos\theta &  -is_2\sin\theta \\
-\sin\theta\cos\theta & \cos^2\theta & is_2 \cos\theta \\ -is_2 \sin\theta &  is_2\cos\theta & -1 \end{array}\right), \nn\\[2ex]
&\qquad\qquad\qquad \ep^{(s_3)}(\q_3)  = \frac{1}{\sqrt{2}}\left(\begin{array}{ccc} \sin^2\phi & -\sin\phi\cos\phi & -is_3\sin\phi \\
-\sin\phi\cos\phi & \cos^2\phi &  is_3 \cos\phi \\ -is_3 \sin\phi & is_3\cos\phi & -1 \end{array}\right).
\end{align}
Here, the helicities $s_i$ take values $\pm 1$, and our conventions for $\ep^{(s_1)}_{ij}(\vec{q}_1)$ are those of \cite{WeinbergBook} (see p.~233).
Note the helicity matrices satisfy the standard identities
\[
\label{PiTTdecomp}
\Pi_{ijkl}(\q) = \half\ep^{(s)}_{ij}(\q)\ep^{(s)}_{kl}(-\q), \qquad
 \ep^{(s)}_{ij}(\q)\ep^{(s')}_{ij}(-\q) = 2\delta^{ss'}.
\]
Defining
\[
S_1 = -q_1^2 + (s_2q_2+s_3q_3)^2, \quad
S_2 = -q_2^2 + (s_3q_3+s_1q_1)^2, \quad
S_3 = -q_3^2 + (s_1q_1+s_2q_2)^2,
\]
we then find
\begin{align}
\label{theta_results}
\theta^{(s_3)}(q_i) & = \frac{\lambda^2}{4\sqrt{2} q_3^2}, \qquad\qquad\qquad\qquad \theta^{(s_2s_3)}(q_i) = \frac{1}{8q_2^2q_3^2}S_1^2, \nn\\[1ex]
\theta^{(s_1s_2s_3)}(q_i) & = \frac{\lambda^2}{32\sqrt{2}q_1^2q_2^2q_3^2}\,(S_1+S_2+S_3)^2
= \frac{\lambda^2}{32\sqrt{2}q_1^2q_2^2q_3^2}\,(s_1q_1+s_2q_2+s_3q_3)^4,
\end{align}
and similarly,
\begin{align}
\label{Theta_results}
&\Theta_1^{(s_3)}(q_i) = -\frac{\lambda^2}{4\sqrt{2} q_1^2 q_3^2}, \qquad
&&\Theta_2^{(s_3)}(q_i) = -\frac{\lambda^2}{4\sqrt{2} q_2^2 q_3^2},
\nn \\[1ex]
&\Theta^{(s_2s_3)}(q_i)  =\frac{1}{8q_2^2q_3^2}S_1^2-\frac{\lambda^2}{16q_1^2q_2^2q_3^2}S_1, \qquad
&&\Theta^{(s_1s_2s_3)}(q_i) = -\frac{1}{16\sqrt{2}q_1^2q_2^2q_3^2}S_1S_2S_3.
\end{align}

\section{Notation for correlators and integration measures} 
\label{App_Conventions}

In this section we collect together various notational devices 
we use throughout the main text.

Firstly, the measures appearing in momentum space integrals are defined as
\begin{align}
\label{shorthand}
&[\d q] = (2\pi)^{-3}\d^3 \vec{q}, \qquad [[\d q_2 \d q_3]] = (2\pi)^3\delta(\sum_i \q_i)[\d q_2][\d q_3], \nn\\[1ex]
&\qquad  [[\d q_1\d q_2\d q_3]] = (2\pi)^3\delta(\sum_i \q_i)[\d q_1][\d q_2][\d q_3].
\end{align}

Secondly, we use a double bracket notation for correlators designed to suppress the appearance of delta functions associated with overall momentum conservation in our formulae.
For cosmological correlators, we define
\begin{align}
\<\z(z,\q)\z(z,\q')\>\ &= (2\pi)^3\delta(\q+\q')\<\!\<\z(z,q)\z(z,-q)\>\!\>, \nn\\
\<\bg^{(s)}(z,\q)\bg^{(s')}(z,\q')\> &= (2\pi)^3\delta(\q+\q')\<\!\<\bg^{(s)}(z,q)\bg^{(s')}(z,-q)\>\!\>, \nn\\
\<\z(z,q_1)\z(z,q_2)\bg^{(s_3)}(z,q_3)\> &= (2\pi)^2\delta(\sum \q_i)\, \<\!\<\z(z,\q_1)\z(z,\q_2)\bg^{(s_3)}(z,\q_3)\>\!\>,
\end{align}
and similarly for stress-energy tensor correlators,
\begin{align}
 \<T_{ij}(\vbq_1)T_{kl}(\vbq_2)\> &= (2\pi)^3\delta(\vbq_1+\vbq_2)\<\!\<T_{ij}(\bq_1)T_{kl}(-\bq_1)\>\!\>, \nn\\
\<T_{ij}(\vbq_1)T_{kl}(\vbq_2)T_{mn}(\vbq_3)\> &= (2\pi)^3\delta(\sum\vbq_i)\<\!\<T_{ij}(\bq_1)T_{kl}(\bq_2)T_{mn}(\bq_3)\>\!\>,  \nn\\
\<T_{ij}(\vbq_1)\Upsilon_{klmn}(\vbq_2,\vbq_3)\> &= (2\pi)^3\delta(\sum\vbq_i)\<\!\<T_{ij}(\bq_1)\Upsilon_{klmn}(\bq_2,\bq_3)\>\!\>.
\end{align}

Finally, it is useful to have a shorthand notation for the various components of the stress-energy tensor 3-point function.
For this, we write
\begin{align}
 \<\!\<T(\bq_1)T(\bq_2)T(\bq_3)\>\!\> &= \delta_{ij}\delta_{kl}\delta_{mn}\<\!\<T_{ij}(\bq_1)T_{kl}(\bq_2)T_{mn}(\bq_3)\>\!\>, \nn\\
 \<\!\<T(\bq_1)T(\bq_2)T^{(s_3)}(\bq_3)\>\!\> &= \half\delta_{ij}\delta_{kl}\ep^{(s_3)}_{mn}(-\vbq_3)\<\!\<T_{ij}(\bq_1)T_{kl}(\bq_2)T_{mn}(\bq_3)\>\!\>,
 \nn\\
 \<\!\<T(\bq_1)T^{(s_2)}(\bq_2)T^{(s_3)}(\bq_3)\>\!\> &=
\frac{1}{4}\delta_{ij}\ep^{(s_2)}_{kl}(-\vbq_2)\ep^{(s_3)}_{mn}(-\vbq_3)\<\!\<T_{ij}(\bq_1)T_{kl}(\bq_2)T_{mn}(\bq_3)\>\!\>, \nn\\
 \<\!\<T^{(s_1)}(\bq_1)T^{(s_2)}(\bq_2)T^{(s_3)}(\bq_3)\>\!\> &=
\frac{1}{8}\ep^{(s_1)}_{ij}(-\vbq_1)\ep^{(s_2)}_{kl}(-\vbq_2)\ep^{(s_3)}_{mn}(-\vbq_3)\<\!\<T_{ij}(\bq_1)T_{kl}(\bq_2)T_{mn}(\bq_3)\>\!\>, \nn\\
\end{align}
while similarly for the semi-local terms
\begin{align}
 \<\!\<T(\bq_1)\Upsilon(\bq_2,\bq_3)\>\!\> &=\delta_{ij}\delta_{kl}\delta_{mn}\<\!\<T_{ij}(\bq_1)\Upsilon_{klmn}(\bq_2,\bq_3)\>\!\> \nn\\
\<\!\<T(\bq_1)\Upsilon^{(s_3)}(\bq_2,\bq_3)\>\!\> &=\half\delta_{ij}\delta_{kl}\ep^{(s_3)}_{mn}(-\vbq_3)\<\!\<T_{ij}(\bq_1)\Upsilon_{klmn}(\bq_2,\bq_3)\>\!\> \nn\\
\<\!\<T(\bq_1)\Upsilon^{(s_3)}(\bq_3,\bq_2)\>\!\> &=\half\delta_{ij}\delta_{kl}\ep^{(s_3)}_{mn}(-\vbq_3)\<\!\<T_{ij}(\bq_1)\Upsilon_{mnkl}(\bq_3,\bq_2)\>\!\> \nn\\
\<\!\<T(\bq_1)\Upsilon^{(s_2s_3)}(\bq_2,\bq_3)\>\!\> &=\frac{1}{4}\delta_{ij}\ep^{(s_2)}_{kl}(-\vbq_2)\ep^{(s_3)}_{mn}(-\vbq_3)\<\!\<T_{ij}(\bq_1)\Upsilon_{klmn}(\bq_2,\bq_3)\>\!\>
\end{align}
and
\begin{align}
 \<\!\<T^{(s_1)}(\bq_1)\Upsilon(\bq_2,\bq_3)\>\!\> &=\half\ep^{(s_1)}_{ij}(-\vbq_1)\delta_{kl}\delta_{mn}\<\!\<T_{ij}(\bq_1)\Upsilon_{klmn}(\bq_2,\bq_3)\>\!\> \nn\\
\<\!\<T^{(s_1)}(\bq_1)\Upsilon^{(s_3)}(\bq_2,\bq_3)\>\!\> &=\frac{1}{4}\ep^{(s_1)}_{ij}(\vbq_1)\delta_{kl}\ep^{(s_3)}_{mn}(-\vbq_3)\<\!\<T_{ij}(\bq_1)\Upsilon_{klmn}(\bq_2,\bq_3)\>\!\> \nn\\
\<\!\<T^{(s_1)}(\bq_1)\Upsilon^{(s_3)}(\bq_3,\bq_2)\>\!\> &=\frac{1}{4}\ep^{(s_1)}_{ij}(\vbq_1)\delta_{kl}\ep^{(s_3)}_{mn}(-\vbq_3)\<\!\<T_{ij}(\bq_1)\Upsilon_{mnkl}(\bq_3,\bq_2)\>\!\> \nn\\
\<\!\<T^{(s_1)}(\bq_1)\Upsilon^{(s_2s_3)}(\bq_2,\bq_3)\>\!\> &=\frac{1}{8}\ep^{(s_1)}_{ij}(-\vbq_1)\ep^{(s_2)}_{kl}(-\vbq_2)\ep^{(s_3)}_{mn}(-\vbq_3)\<\!\<T_{ij}(\bq_1)\Upsilon_{klmn}(\bq_2,\bq_3)\>\!\> .
\end{align}

Note in particular that, from \eqref{Upsilon_def}, correlators involving $\Upsilon(\bq_2,\bq_3)$ and $\Upsilon^{(s_2s_3)}(\bq_2,\bq_3)$ are symmetric under exchange of $\bq_2$ and $\bq_3$,
\begin{align}
\<\!\<T(\bq_1)\Upsilon(\bq_2,\bq_3)\>\!\>  &=  \<\!\<T(\bq_1)\Upsilon(\bq_3,\bq_2)\>\!\>, \nn\\
\<\!\<T^{(s_1)}(\bq_1)\Upsilon(\bq_2,\bq_3)\>\!\> &=  \<\!\<T^{(s_1)}(\bq_1)\Upsilon(\bq_3,\bq_2)\>\!\>, \nn\\
\<\!\<T(\bq_1)\Upsilon^{(s_2s_3)}(\bq_2,\bq_3)\>\!\> &= \<\!\<T(\bq_1)\Upsilon^{(s_3s_2)}(\bq_3,\bq_2)\>\!\>,\nn\\
\<\!\<T^{(s_1)}(\bq_1)\Upsilon^{(s_2s_3)}(\bq_2,\bq_3)\>\!\> &=\<\!\<T^{(s_1)}(\bq_1)\Upsilon^{(s_3s_2)}(\bq_3,\bq_2)\>\!\> ,
\end{align}
whereas those involving $\Upsilon^{(s_3)}$ are not:
\begin{align}
  \<\!\<T(\bq_1)\Upsilon^{(s_3)}(\bq_2,\bq_3)\>\!\> &= \<\!\<T(\bq_1)\Upsilon^{(s_3)}(\bq_3,\bq_2)\>\!\> - \frac{3}{8}\Theta_1^{(s_3)}(\bq_i)\<\!\<T(\bq_1)T(-\bq_1)\>\!\>, \nn\\
\<\!\<T^{(s_1)}(\bq_1)\Upsilon^{(s_3)}(\bq_2,\bq_3)\>\!\> &= \<\!\<T^{(s_1)}(\bq_1)\Upsilon^{(s_3)}(\bq_3,\bq_2)\>\!\>-\frac{3}{8}\theta^{(s_1s_3)}(\bq_i)A(\bq_1).
\end{align}
In these equations, $\theta^{(s_2s_3)}(\bq_i)$ and $\Theta_1^{(s_3)}(\bq_i)$ are as defined in \eqref{theta_def},
while $A(\bq)$ is the transverse traceless part of the stress-energy tensor 2-point function:
\[
\label{ABdef}
 \<\!\<T_{ij}(\bq)T_{kl}(-\bq)\>\!\> = A(\bq)\Pi_{ijkl}+B(\bq)\pi_{ij}\pi_{kl}.
\]
From this standard result it follows that
\begin{align}
 \<\!\<T(\bq)T(-\bq)\>\!\>&=\delta_{ij}\delta_{kl}\<\!\<T_{ij}(\bq)T_{kl}(-\bq)\>\!\> = 4B(\bq), \nn\\
\<\!\<T^{(s)}(\bq)T(-\bq)\>\!\>&=\half\ep^{(s)}_{ij}(-\vbq)\delta_{kl}\<\!\<T_{ij}(\bq)T_{kl}(-\bq)\>\!\> = 0, \nn\\
\<\!\<T^{(s)}(\bq)T^{(s')}(-\bq)\>\!\>&=\frac{1}{4}\ep^{(s)}_{ij}(-\vbq)\ep^{(s')}_{kl}(\vbq)\<\!\<T_{ij}(\bq)T_{kl}(-\bq)\>\!\> = \half A(\bq)\,\delta^{ss'}.
\end{align}

\section{Constraint equations}

\label{App_constraints}

In this appendix, we present the domain-wall Hamiltonian and momentum constraint equations to quadratic order, as required for our holographic calculations.  We work in synchronous (Fefferman-Graham) gauge where $N_i=0$ and $N=1$.  
(For results including a non-zero lapse perturbation as required for the asymptotically power-law case please see \cite{McFadden:2010vh}).

The full Hamiltonian constraint reads
\[
 0=-R + K^2 -K_{ij}K^{ij}+2\bK^2 V -N^{-2}\dot{\Phi}^2+g^{ij}\Phi_{,i}\Phi_{,j},
\]
where $K_{ij}=(1/2N)\dot{g}_{ij}$ is the extrinsic curvature of constant-$z$ slices.
Expanding to quadratic order, we find
\begin{align}
\label{Ham_constr}
 0 &= -4a^{-2}\D^2\psi+2H\dot{h} -2\dot{\vphi}\delta\dot{\vphi}+2\bK^2 V' \delta\vphi \nn \\
&\quad  -R_{(2)} + \frac{1}{4}\dot{h}^2-\frac{1}{4}\dot{h}_{ij}\dot{h}_{ij}-2Hh_{ij}\dot{h}_{ij} 
 -\delta\dot{\vphi}^2 +\bK^2 V'' \delta\vphi^2+a^{-2}\delta\vphi_{,i}\delta\vphi_{,i},
\end{align}
where repeated covariant indices are to be summed over using the Kronecker delta, and $h \equiv h_{ii}$.
For the purposes of our holographic calculations, we will not need to evaluate $R_{(2)}$ explicitly.

Similarly, the momentum constraint
\[
 0 = \nabla_j (K^j_i -\delta^j_i K)-N^{-1}\dot{\Phi}\Phi_{,i},
\]
yields
\[
 0 = \dot{h}_{ij,j}-\dot{h}_{,i} -2\dot{\vphi}\delta\vphi_{,i} 
 +\frac{1}{2}h_{,j}\dot{h}_{ji}-\frac{1}{2}\dot{h}_{jk} h_{jk,i} -2\delta\dot{\vphi}\delta\vphi_{,i} 
 + (h_{jk}\dot{h}_{jk})_{,i} -  (h_{jk}\dot{h}_{ki})_{,j}
\]
when expanded to quadratic order.
Extracting the scalar part by acting with $\D^{-2}\D_i$, we find
\[
\label{mom_constr}
 0 = 2\dot{\psi}-\dot{\vphi}\delta\vphi + \half h_{jk}\dot{h}_{jk} 
+\D^{-2}\D_i\Big[ \frac{1}{4}(h_{,j}\dot{h}_{ji}-\dot{h}_{jk}h_{jk,i})-\half (h_{jk}\dot{h}_{ki})_{,j}-\delta\dot{\vphi}\delta\vphi_{,i}\Big].
\]

\section{Holographic analysis continued}
\label{App_hol}

Here we present the remaining part of the holographic analysis not covered in Section \ref{Hol_section}.  
The quantities $\Theta_1^{(s_3)}$, $\Theta_2^{(s_3)}$, $\Theta^{(s_2s_3)}$ and $\Theta^{(s_2s_3s_3)}$, as well as
 $\theta^{(s_3)}$, $\theta^{(s_2s_3)}$ and $\theta^{(s_2s_3s_3)}$ are defined in Appendix \ref{App_helicity}.
Our conventions for correlators and momentum space integration measures are given in Appendix \ref{App_Conventions}.

\subsection{Cross-check for $\<TTT^{(s)}\>$}

In this subsection, we expand $\delta\<T^{(s_3)}(\vbq_3)\>_s$ to quadratic order and compute the coefficient of the $\psi(-\vbq_2)\psi(-\vbq_3)$ term.
This calculation serves as a useful and nontrivial cross-check of our earlier result \eqref{cross_check3} for $\<\!\< T(\bq_1)T(\bq_2)T^{(s_3)}(\bq_3)\>\!\>$.  

We start by setting the sources to $h_{ij}=-2\psi\delta_{ij}$ and $\delta\vphi=0$.
From \eqref{1ptfnTT}, using \eqref{rules}, we then have
\begin{align}
\label{gzz_eq0}
\delta\<T^{(s_3)}(\vbq_3)\>_s =  \bK^{-2}\Big[-\half\dot{\g}^{(s_3)}(\vbq_3) 
+ \int[[\d\bq_1\d\bq_2]]
\Big[ & \frac{1}{4a^3}\Big(\Theta_1^{(s_3)}(\bq_i)\bOi(\bq_1)+\Theta_2^{(s_3)}(\bq_i)\bOi(\bq_2)\Big) \nn\\
& +\frac{1}{a^2H}\theta^{(s_3)}(\bq_i) \Big]
\psi(-\vbq_1)\psi(-\vbq_2)+\ldots\Big]_{(3)}.
\end{align}
From the definition \eqref{gb_gi} of the gauge-invariant variable $\bg_{ij}$, in synchronous gauge we have
\begin{align}
 \bg_{ij} &= \g_{ij} +\ldots \\
\dot{\bg}_{ij}&= \dot{\g}_{ij}+\Pi_{ijkl}\Big[-\frac{\delta\dot{\vphi}}{\dot{\vphi}}\dot{h}_{kl}-2(\dot{\chi}_{,mk}+\dot{\omega}_{m,k})(-2\psi\delta_{ml})+2\psi\dot{\g}_{kl}\Big]+\ldots
\end{align}
where we have omitted terms that vanish when the sources are restricted to $h_{ij}=-2\psi\delta_{ij}$ and $\delta\vphi=0$.
Applying \eqref{rules} to the quadratic terms, we find
\begin{align}
\label{gzz_eq1}
\bg^{(s_3)}(\vbq_3) &= \g^{(s_3)}(\vbq_3) + \ldots \\[1ex]
\label{gzz_eq2}
 \dot{\bg}^{(s_3)}(\vbq_3) &=\dot{\g}^{(s_3)}(\vbq_3) + \int[[\d\bq_1\d\bq_2]]
\Big[ \frac{1}{8a^6\ep H}\Big(\Theta_1^{(s_3)}(\bq_i)+\Theta_2^{(s_3)}(\bq_i)\Big)\bOi(\bq_1)\bOi(\bq_2) \nn\\[1ex]
&\qquad +\frac{1}{4a^5\ep H^2}\theta^{(s_3)}(\bq_i)\big(\bOi(\bq_1)+\bOi(\bq_2)\big)-\frac{1}{2a^3}\Big(\Theta_1^{(s_3)}(\bq_i)\bOi(\bq_1)+\Theta_2^{(s_3)}(\bq_i)\bOi(\bq_2)\Big)
\nn\\[1ex]
&\qquad -\frac{2}{a^2H}\theta^{(s_3)}(\bq_i)\Big] \times \psi(-\vbq_1)\psi(-\vbq_2)+\ldots
\end{align}
On the other hand, from Hamilton's equations \eqref{Hamilton_eqs}, we have
\begin{align}
\label{gzz_eq3}
 \dot{\bg}^{(s_3)}(\vbq_3) &=\frac{4}{a^3}\bEi(\bq_3)\g^{(s_3)}(\vbq_3)+\int[[\d\bq_1\d\bq_2]]
\Big[\frac{4}{a^3}\bEii^{(s_3)}(\bq_3,\bq_1\bq_2)+\half\mB^{(s_3)}_{123}+\frac{1}{4}\mD^{(s_3)}_{123}\bOi(\bq_2) \nn\\[1ex]
& \qquad\qquad +\frac{1}{4}\mD^{(s_3)}_{213}\bOi(\bq_1)
+\half\mF^{(s_3)}_{123}\bOi(\bq_1)\bOi(\bq_2)\Big]\psi(-\vbq_1)\psi(-\vbq_2)+\ldots
\end{align}
where we used \eqref{gzz_eq1} and the fact that $\z(-\vbq_1)\z(-\vbq_2)=\psi(-\vbq_1)\psi(-\vbq_2)$ with the source $\delta\vphi$ set to zero.
Solving \eqref{gzz_eq2} and \eqref{gzz_eq3} for $\dot{\g}^{(s_3)}(\vbq_3)$, and inserting into \eqref{gzz_eq0}, after cancellations we obtain
\begin{align}
 \delta\<T^{(s_3)}(\vbq_3)\>_s &=
 \bK^{-2}\Big[-\frac{2}{a^3}\bEi(\bq_3)\g^{(s_3)}(\vbq_3)+\int[[\d\bq_1\d\bq_2]]\Big[-\frac{2}{a^3}\bEii^{(s_3)}(\bq_3,\bq_1,\bq_2) \nn\\[1ex]
& +\frac{3}{8a^3}\Big(\Theta_1^{(s_3)}(\bq_i)\bOi(\bq_1)+\Theta_2^{(s_3)}(\bq_i)\bOi(\bq_2)\Big)+\frac{3}{2a^2H}\theta^{(s_3)}(\bq_i)\Big]\psi(-\vbq_1)\psi(-\vbq_2)+\ldots\Big]_{(3)}.
\end{align}
Here, the last term is an ultralocal contact term (it is finite under sending any of the $\bq_i$ to zero) and so we discard it.
We therefore obtain
\begin{align}
  \delta\<T^{(s_3)}(\vbq_3)\>_s &= -2\bK^{-2}\bar{E}_{[2](0)}(\bq_3)\g^{(s_3)}_{(0)}(\vbq_3) +
\int[[\d\bq_1\d\bq_2]]\Big[-2\bK^{-2}\bar{E}_{[3](0)}^{(s_3)}(\bq_3,\bq_1,\bq_2) \nn\\[1ex] &\qquad +\frac{3}{8}\bK^{-2}\Big(\Theta_1^{(s_3)}(\bq_i)\bOi(\bq_1)+\Theta_2^{(s_3)}(\bq_i)\bOi(\bq_2)\Big)\Big]\psi_{(0)}(-\vbq_1)\psi_{(0)}(-\vbq_2)+\ldots
\end{align}
Comparing with \eqref{1ptTTexp} (after a suitable permutation), we recover \eqref{E2_result} and the result
\begin{align}
\label{cross_check2}
-4\bK^{-2}\bar{E}_{[3](0)}^{(s_3)}(\bq_3,\bq_1,\bq_2) &=
\<\!\<T(\bq_1)T(\bq_2)T^{(s_3)}(\bq_3)\>\!\>
-\frac{1}{2}\Theta_1^{(s_3)}(\bq_i)\<\!\<T(\bq_1)T(-\bq_1)\>\!\>  \nn\\
&\quad -\frac{1}{2}\Theta_2^{(s_3)}(\bq_i)\<\!\<T(\bq_2)T(-\bq_2)\>\!\>
 -2\<\!\<\Upsilon(\bq_1,\bq_2)T^{(s_3)}(\bq_3)\>\!\> \nn\\[1ex]
&\quad -2\<\!\<T(\bq_1)\Upsilon^{(s_3)}(\bq_2,\bq_3)\>\!\>
-2\<\!\<T(\bq_2)\Upsilon^{(s_3)}(\bq_1,\bq_3)\>\!\>.
\end{align}
From \eqref{cross_relations}, we see that this result agrees perfectly with \eqref{cross_check1} in Section \ref{Hol_section}.

\subsection{Computation of $\<T T^{(s)}T^{(s)}\>$}

As noted in Section \ref{Hol_section}, there are two ways of calculating this 3-point function.  The first method is to expand $\delta\<T^i_i(\vbq_1)\>_s$ to quadratic order in $\g^{(s_2)}(-\vbq_2)\g^{(s_3)}(-\vbq_3)$, while the second is to expand $\delta\<T^{(s_3)}(\vbq_3)\>_s$ to quadratic order in $\psi(-\vbq_1)\g^{(s_2)}(-\vbq_2)$.  The results from both methods should agree.

\subsubsection*{\it (i) First method}

Here we set the sources to $h_{ij}=\g_{ij}$ and $\delta\vphi=0$.  Examining the momentum constraint, using \eqref{rules} to replace momenta in quadratic terms, we find
\begin{align}
\label{zgg_Ham}
(\dot{h}-h_{ij}\dot{h}_{ij})(\vbq_1) &= \frac{\dot{\vphi}}{H}\delta\dot{\vphi}(\vbq_1)+\int[[\d\bq_2\d\bq_3]]\frac{2}{a^6 H}\bEi(\bq_2)\bEi(\bq_3)\theta^{(s_2s_3)}(\bq_i)\g^{(s_2)}(\vbq_2)\g^{(s_3)}(-\vbq_3)+\ldots
\end{align}
From the definition of $\z$ in \eqref{zeta_gi}, we have
\begin{align}
\label{zexpgg}
 \z(\vbq_1) &= -\frac{1}{8}\int[[\d\bq_2\d\bq_3]]\Theta^{(s_2s_3)}(\bq_i)\g^{(s_2)}(-\vbq_2)\g^{(s_3)}(-\vbq_3)+\ldots, \\[1ex]
\dot{\z}(\vbq_1) &= -\dot{\psi}(\vbq_1)-\frac{H}{\dot{\vphi}}\delta\dot{\vphi}(\vbq_1) \nn\\ & \qquad -\int[[\d\bq_2\d\bq_3]]\frac{1}{2a^3}\big(\bEi(\bq_2)+\bEi(\bq_3)\big)\Theta^{(s_2s_3)}(\bq_i)\g^{(s_2)}(-\vbq_2)\g^{(s_3)}(-\vbq_3)+\ldots
\end{align}
On the other hand, Hamilton's equation \eqref{zdot} combined with \eqref{zexpgg} yields
\begin{align}
\label{zgg_eom}
 \dot{\z}(\vbq_1) &= \int[[\d\bq_2\d\bq_3]]\Big[\frac{1}{2a^3\ep}\bOii^{(s_2s_3)}(\bq_i)-\frac{1}{16a^3\ep}\Theta^{(s_2s_3)}(\bq_i)\bOi(\bq_1)
+\mD^{(s_2s_3)}_{123}+\half\mE^{(s_2s_3)}_{123}\bEi(\bq_3) \nn\\
& \qquad\qquad\qquad\quad +\half\mE^{(s_3s_2)}_{132}\bEi(\bq_2)+\mF^{(s_2s_3)}_{123}\bEi(\bq_2)\bEi(\bq_3)\Big]\g^{(s_2)}(-\vbq_2)\g^{(s_3)}(-\vbq_3) +\ldots
\end{align}
where the coefficients appearing in this equation are defined in \eqref{zgg_coeffts}.

The momentum constraint \eqref{mom_constr}, after applying \eqref{rules} to the quadratic terms, reads
\begin{align}
\label{zgg_mom}
\dot{\psi}(\vbq_1) &= \int[[\d\bq_2\d\bq_3]]\Big[-\frac{1}{2a^3}\big(\bEi(\bq_2)+\bEi(\bq_3)\big)\Theta^{(s_2s_3)}(\bq_i) \nn\\[1ex]
& -\frac{1}{4a^3 \bq_1^2}\Big((\vbq_1\cdot\vbq_2)\bEi(\bq_3)+(\vbq_1\cdot\vbq_3)\bEi(\bq_2)\big)\theta^{(s_2s_3)}(\bq_i)\Big]\g^{(s_2)}(-\vbq_2)\g^{(s_3)}(-\vbq_3)+\ldots
\end{align}
Combining \eqref{zgg_Ham}, \eqref{zexpgg}, \eqref{zgg_eom} and \eqref{zgg_mom}, we find
\begin{align}
 (\dot{h}-h_{ij}\dot{h}_{ij})(\vbq_1) &= \int[[\d\bq_2\d\bq_3]]\Big[-\frac{1}{a^3}\bOii^{(s_2s_3)}(\bq_i)+\frac{1}{8a^3}\Theta^{(s_2s_3)}(\bq_i)\bOi(\bq_1) \nn\\
&\qquad \qquad
-\frac{1}{8a^2H}(\vbq_2\cdot\vbq_3)\theta^{(s_2s_3)}(\bq_i)\Big]\g^{(s_2)}(-\vbq_2)\g^{(s_3)}(-\vbq_3)+\ldots
\end{align}
The last term in this expression is an ultralocal contact term (it is finite upon taking any of the $\bq_i$ to zero) and so may be discarded.
We then find
\[
 \delta\<T^i_i(\vbq_1)\>_s = \int[[\d\bq_2\d\bq_3]]\Big[-\bK^{-2}\bar{\Omega}_{[3](0)}^{(s_2s_3)}(\bq_i)+\frac{1}{8}\bK^{-2}\bar{\Omega}_{[2](0)}(\bq_1)\Theta^{(s_2s_3)}(\bq_i)\Big]
\g^{(s_2)}(-\vbq_2)\g^{(s_3)}(-\vbq_3)+\ldots
\]
Comparing with \eqref{1ptTrexp}, we then recover the result \eqref{cross_check3} given in Section \ref{Hol_section}.

\subsubsection*{\it (ii) Second method}

In this calculation we set the sources to $h_{ij} = -2\psi\delta_{ij}+\g_{ij}$ and $\delta\vphi=0$, and collect all quadratic terms of the form
$\psi(-\vbq_1)\g^{(s_2)}(-\vbq_2)$.

Beginning with \eqref{1ptfnTT} and replacing the quadratic terms with \eqref{rules}, we obtain
\begin{align}
\label{gzg_eq0}
 \delta\<T^{(s_3)}(\vbq_3)\>_s &= \bK^{-2}\Big[-\half\dot{\g}^{(s_3)}(\vbq_3)+\int[[\d\bq_1\d\bq_2]]\Big[
-\frac{2}{a^3}\theta^{(s_2s_3)}(\bq_i)\bEi(\bq_2) \nn\\[1ex] &\qquad +\frac{1}{4}\big(\theta^{(s_2s_3)}(\bq_i)-\Theta^{(s_2s_3)}(\bq_i)\big)\Big(\frac{\bOi(\bq_1)}{a^3}-\frac{2\bq_1^2}{a^2H}\Big)\Big]\psi(-\vbq_1)\g^{(s_2)}(-\vbq_2)+\ldots\Big]_{(3)}
\end{align}
From the definition of $\bg_{ij}$, \eqref{gb_gi}, evaluated in synchronous gauge,
\begin{align}
 \bg_{ij}&=\g_{ij}+\Pi_{ijkl}\big[2\psi\g_{kl}\big] + \ldots, \\
\dot{\bg}_{ij} &= \dot{\g}_{ij}+\Pi_{ijkl}\big[-\frac{\delta\dot{\vphi}}{\dot{\vphi}}\dot{h}_{kl}-2(\dot{\chi}_{,mk}+\dot{\omega}_{m,k})(-2\psi\delta_{ml}+\g_{ml})
-(\dot{\chi}_{,m}+\dot{\omega}_m)(-2\psi_{,m}\delta_{kl}+\g_{kl,m}) \nn\\
&\qquad\qquad\qquad +2\dot{\psi}\g_{kl}+2\psi\dot{\g}_{kl}-\dot{\g}_{km}\g_{ml}\big]+\ldots,
\end{align}
where we omit terms that vanish when setting the sources to $h_{ij} = -2\psi\delta_{ij}+\g_{ij}$ and $\delta\vphi=0$.
Replacing the quadratic terms in momentum space using \eqref{rules}, we obtain
\begin{align}
\label{gzg_eq1}
 \bg^{(s_3)}(\vbq_3) &= \g^{(s_3)}(\vbq_3)+\int[[\d\bq_1\d\bq_2]]\theta^{(s_2s_3)}(\bq_i)\psi(-\vbq_1)\g^{(s_2)}(-\vbq_2)+\ldots \\
\label{gzg_eq2}
\dot{\bg}^{(s_3)}(\vbq_3) &= \dot{\g}^{(s_3)}(\vbq_3)+\int[[\d\bq_1\d\bq_2]]\Big[
\frac{4}{a^3}\bEi(\bq_2)\theta^{(s_2s_3)}(\bq_i)-\frac{1}{a^6\ep H}\theta^{(s_2s_3)}(\bq_i)\bEi(\bq_2)\bOi(\bq_1) \nn\\[1ex]
&\qquad\quad -\frac{1}{4}\Big(\frac{\bOi(\bq_1)}{a^3}-\frac{2\bq_1^2}{a^2H}\Big)\Big(\frac{(\vbq_1\cdot\vbq_2)}{\bq_1^2}\theta^{(s_2s_3)}(\bq_i)+2\big(\theta^{(s_2s_3)}(\bq_i)-\Theta^{(s_2s_3)}(\bq_i)\big)\Big)\Big] \nn\\[1ex]
&\qquad\qquad\qquad\qquad \times\psi(-\vbq_1)\g^{(s_2)}(-\vbq_2) +\ldots
\end{align}
On the other hand, from Hamilton's equations \eqref{Hamilton_eqs}, along with \eqref{response_fns1}, we have
\begin{align}
\label{gzg_eq3}
 \dot{\bg}^{(s_3)}(\vbq_3)
&= \frac{4}{a^3}\bEi(\bq_3)\g^{(s_3)}(\vbq_3)+\int[[\d\bq_1\d\bq_2]]\Big[-\half\mB^{(s_2s_3)}_{123} -\mC^{(s_2s_3)}_{123}\bEi(\bq_2)
-\half\mE^{(s_2s_3)}_{123}\bOi(\bq_1) \nn\\[1ex]
&\qquad -\mF^{(s_2s_3)}_{123}\bOi(\bq_1)\bEi(\bq_2)-\frac{4}{a^3}\bEii^{(s_3s_2)}(\bq_3,\bq_1,\bq_2)+\frac{4}{a^3}\theta^{(s_2s_3)}(\bq_i)\bEi(\bq_3)\Big]
\nn\\[1ex]
&\qquad\qquad\qquad\quad \times \psi(-\vbq_1)\g^{(s_2)}(-\vbq_2)+\ldots
\end{align}
where we also used \eqref{gzg_eq1} and the fact that $\z(-\vbq_1)\bg^{(s_2)}(-\vbq_2)=-\psi(-\vbq_2)\g^{(s_2)}(-\vbq_2)$ when $\delta\vphi=0$.
Solving \eqref{gzg_eq2} and \eqref{gzg_eq3} for $\dot{\g}^{(s_3)}(\vbq_3)$ and backsubstituting into \eqref{gzg_eq0}, we find
\begin{align}
 \delta\<T^{(s_3)}(\vbq_3)\>_s & =-2\bK^{-2}\bar{E}_{[2](0)}(\bq_3)\g^{(s_3)}(\vbq_3)+\int[[\d\bq_1\d\bq_2]]\Big[2\bK^{-2}\bar{E}^{(s_3s_2)}_{[3](0)}(\bq_3,\bq_1,\bq_2) \nn\\[1ex]
&\qquad
-\theta^{(s_2s_3)}(\bq_i)\bK^{-2}\big(2\bar{E}_{[2](0)}(\bq_3)+3\bar{E}_{[2](0)}(\bq_2)\big)\Big]\psi_{(0)}(-\vbq_1)\g_{(0)}^{(s_2)}(-\vbq_2)+\ldots
\end{align}
Comparing with \eqref{1ptTTexp}, we recover \eqref{E2_result} and find
\begin{align}
\label{cross_check4}
-\bK^{-2}\bar{E}^{(s_3s_2)}_{[3](0)}(\bq_3,\bq_1,\bq_2) &=
 \half\<\!\<T(\bq_1)T^{(s_2)}(\bq_2)T^{(s_3)}(\bq_3)\>\!\>
-\frac{1}{4}\big(A(\bq_2)+A(\bq_3)\big)\theta^{(s_2s_3)}(\bq_i) \nn\\[1ex]
&\quad -\frac{1}{8}\<\!\<T(\bq_1)T(-\bq_1)\>\!\>\Theta^{(s_2s_3)}(\bq_i)
-\<\!\<T(\bq_1)\Upsilon^{(s_2s_3)}(\bq_2,\bq_3)\>\!\> \nn\\[1ex]
&\quad
-\<\!\<T^{(s_2)}(\bq_2)\Upsilon^{(s_3)}(\bq_1,\bq_3)\>\!\>
-\<\!\<T^{(s_3)}(\bq_3)\Upsilon^{(s_2)}(\bq_1,\bq_2)\>\!\> .
\end{align}
From \eqref{cross_relations}, this result agrees with \eqref{cross_check3} from the previous subsection.
The two calculations are therefore consistent providing once again a useful and nontrivial cross-check.

\subsection{Computation of $\<T^{(s)}T^{(s)}T^{(s)}\>$}

Here we need to expand $\delta\<T^{(s_1)}\>_s$ to quadratic order and obtain the coefficient of the $\g^{(s_2)}\g^{(s_3)}$ term.
We will therefore set the sources to $h_{ij}=\g_{ij}$ and $\delta\vphi=0$ in the following.
From \eqref{1ptfnTT} and \eqref{rules}, we have
\begin{align}
\delta\<T^{(s_1)}(\vbq_1)\>_s
= \bK^{-2}\Big[-\half\dot{\g}^{(s_1)}(\vbq_1) 
 + &\int[[\d\bq_2\d\bq_3]]\frac{1}{2a^3} \big(\bEi(\bq_2)+\bEi(\bq_3)\big) \nn\\
&\qquad \times \Theta^{(s_1s_2s_3)}(\bq_i)\g^{(s_2)}(-\vbq_2)\g^{(s_3)}(-\vbq_3)+\ldots\Big]_{(3)}.
\end{align}
From the definition \eqref{gb_gi} of the gauge-invariant variable $\bg_{ij}$, however, in synchronous gauge we have
\begin{align}
\label{eq5}
\bg^{(s_1)}(\vbq_1) &= \g^{(s_1)}(\vbq_1)-\frac{1}{4}\int[[\d\bq_2\d\bq_3]]\Theta^{(s_1s_2s_3)}(\bq_i)\g^{(s_2)}(-\vbq_2)\g^{(s_3)}(-\vbq_3)+\ldots, \\[1ex]
\label{eq6}
\dot{\bg}^{(s_1)}(\vbq_1) &=\dot{\g}^{(s_1)}(\vbq_1)-\int[[\d\bq_2\d\bq_3]]\frac{1}{a^3}\big(\bEi(\bq_2)+\bEi(\bq_3)\big)\Theta^{(s_1s_2s_3)}(\bq_i)\g^{(s_2)}(-\vbq_2)\g^{(s_3)}(-\bq_3)+\ldots
\end{align}
where in the last line we used \eqref{rules} to replace the $\dot{\g}$ in the quadratic term.

Hamilton's equations \eqref{Hamilton_eqs}, when combined with \eqref{response_fns1}, give us
\begin{align}
\dot{\bg}^{(s_1)}(\vbq_1)
& =\frac{4}{a^3}\bEi(\bq_1)\g^{(s_1)}(\vbq_1) \nn\\[1ex]
&\qquad +\int[[\d\bq_2\d\bq_3]]\frac{1}{a^3}\Big[4\bEii^{(s_1s_2s_3)}(\bq_i)-\bEi(\bq_1)\Theta^{(s_1s_2s_3)}(\bq_i)\Big]
\g^{(s_2)}(-\vbq_2)\g^{(s_3)}(-\vbq_3)+\ldots
\end{align}
where we additionally made use of 
\eqref{eq5}.
Equating with \eqref{eq6} we may solve for $\dot{\g}^{(s_1)}(\vbq_1)$, whence
\begin{align}
\delta\<T^{(s_1)}(\vbq_1)\>_s &= -2\bK^{-2}\bar{E}_{[2](0)}(\bq_1)\g^{(s_1)}_{(0)}(\vbq_1) 
+\int[[\d\bq_2\d\bq_3]]\Big[-2\bK^{-2}\bar{E}_{[3](0)}^{(s_1s_2s_3)}(\bq_i) \nn\\[1ex] & \qquad\qquad +\half\bK^{-2}\bar{E}_{[2](0)}(\bq_1)\Theta^{(s_1s_2s_3)}(\bq_i)\Big] \g^{(s_2)}_{(0)}(-\vbq_2)\g^{(s_3)}_{(0)}(-\vbq_3) 
+\ldots
\end{align}
Finally, comparing with the relevant portion of \eqref{1ptTTexp}, we recover \eqref{E2_result} and the result
\eqref{Esss_result} presented in Section \ref{Hol_section}.


\end{document}